\documentclass[nonacm]{acmart}

\usepackage[T1]{fontenc}
\usepackage[utf8]{inputenc}
\usepackage{amsfonts}
\usepackage{fancyhdr}
\usepackage{appendix}
\usepackage{float}
\usepackage{hyperref}

\usepackage{subcaption}
\usepackage{multirow}

\usepackage{enumitem}
\setlist{leftmargin=*,itemsep=0pt}

\usepackage{centernot}
\usepackage[linesnumbered,ruled,vlined]{algorithm2e}

\usepackage{color}
\usepackage{soul}

\definecolor{lightgrey}{rgb}{0.925, 0.925, 0.925}
\sethlcolor{lightgrey}

\makeatletter
\def\SOUL@hlpreamble{%
    \setul{}{3.5ex}
    \let\SOUL@stcolor\SOUL@hlcolor
    \dimen@\SOUL@ulthickness
    \dimen@i=-.75ex 
    \advance\dimen@i-.5\dimen@
    \edef\SOUL@uldepth{\the\dimen@i}%
    \let\SOUL@ulcolor\SOUL@stcolor
    \SOUL@ulpreamble
}
\makeatother

\newcommand*{\codebox}[1]{{\ttfamily\hyphenchar\font=45\relax\hl{~#1~}}}

\usepackage{listings}

\SetCommentSty{mycommfont}
\definecolor{cppColorBackground}{rgb}{1.,1.,1.}
\definecolor{cppColorComment}{rgb}{0.0,0.47,.8}
\definecolor{cppColorLine}{rgb}{0.6,0.6,0.6}
\definecolor{cppColorString}{rgb}{0,0.501,145}
\definecolor{cppColorKey}{rgb}{0.8,0.5,0}
\definecolor{cppColorDigit}{rgb}{0,0,0.5}

\usepackage{placeins} 

\usepackage{pgfplots}
\pgfplotsset{compat=newest}
\usetikzlibrary{patterns}
\usepackage{pgfplotstable}

\AtBeginDocument{%
  \providecommand\BibTeX{{%
    \normalfont B\kern-0.5em{\scshape i\kern-0.25em b}\kern-0.8em\TeX}}}
    
\begin{document}

\title[Autovesk: Automatic vectorized code generation]{Autovesk: Automatic vectorized code generation from unstructured static kernels using graph transformations}

\author{Hayfa Tayeb}
\affiliation{%
  \institution{ICube Lab}
  \streetaddress{300 Bd Sébastien Brant}
  \city{Illkirch}
  \country{France}
}
\affiliation{%
  \institution{Inria}
  \city{Nancy}
  \country{France}
}
\affiliation{%
  \institution{University of Strasbourg}
  \country{France}
}
\email{hayfa.tayeb@inria.fr}

\author{Ludovic Paillat}
\affiliation{%
  \institution{ICube Lab}
  \streetaddress{300 Bd Sébastien Brant}
  \city{Illkirch}
  \country{France}
}
\affiliation{%
  \institution{Inria}
  \city{Nancy}
  \country{France}
}
\affiliation{%
  \institution{University of Strasbourg}
  \country{France}
}
\email{ludovic.paillat@etu.unistra.fr}

\author{B\'erenger Bramas}
\affiliation{%
  \institution{ICube Lab}
  \streetaddress{300 Bd Sébastien Brant}
  \city{Illkirch}
  \country{France}
}
\affiliation{%
  \institution{Inria}
  \city{Nancy}
  \country{France}
}
\affiliation{%
  \institution{University of Strasbourg}
  \country{France}
}
\email{berenger.bramas@inria.fr}

\begin{abstract}
Leveraging the SIMD capability of modern CPU architectures is mandatory to take full advantage of their increased performance.
To exploit this capability, binary executables must be vectorized, either manually by developers or automatically by a tool.
For this reason, the compilation research community has created several strategies to transform a scalar code into a vectorized implementation.
However, most existing automatic vectorization techniques in modern compilers are designed for regular codes, leaving irregular applications with non-contiguous data access patterns at a disadvantage. 
We present in this paper a new tool, Autovesk, that automatically generates vectorized code from scalar code, specifically targeting irregular data access patterns.  
We describe how our method transforms a graph of scalar instructions into a vectorized one using different heuristics to reduce the number or cost of the instructions.
Finally, we demonstrate the effectiveness of our approach on various computational kernels using Intel AVX-512 and ARM SVE.
We compare the speedups of Autovesk vectorized code over GCC, Clang LLVM and Intel automatic vectorization optimizations.
We achieve competitive results in linear kernels and up to 11x speedups in irregular ones.
\end{abstract}



\maketitle

\section{Introduction}

Vectorization is a feature of modern CPUs that consists in applying a single instruction to multiple data (SIMD)~\cite{5009071}.
It is one of the hardware features that have increased CPU performance despite the stagnation of the clock frequency.
To exploit this functionality, a binary must explicitly use vector instructions, and this requires the program to be either vectorized by the back-end compiler when it issues the machine code, or at runtime~\cite{hallou:hal-01593216}.

HPC experts tend to vectorize their code manually~\cite{10.7717/peerj-cs.151} in assembly or with intrinsic functions~\footnote{An intrinsic function is a function that is usually converted into a single instruction by the compiler, allowing to use some instructions explicitly while remaining at a high level}.
This allows them to fine-tune and control the behavior of their computational kernels.
However, this also requires significant expertise and can imply re-implement new kernels for each instruction set architecture (ISA).
To mitigate this drawback, several abstraction libraries have been created~\cite{bramas2017inastemp,falcou2004application,10.1145/2568058.2568059,7568423,10.1145/2568058.2568062,souza2015openvec,moller2016design}.
They allow writing vectorized code with an abstract vector type which is fixed at compile time depending on the target architecture.
This clearly facilitates maintainability while still asking for a moderate understanding of the vectorization concept.
The main limitation of this approach is that the operations supported by the library should exist in all underlying ISA or, when it is not the case, the desired behavior should be implemented with more instructions, leading to difficulty to have a single kernel implementation efficient everywhere.
From another research perspective, the compilation community proposed different strategies for automatically transforming a scalar code into a vectorized equivalent. 
The resulting mechanisms introduced in modern compilers work well when the code is regular, i.e. with contiguous data accesses and affine loops. 
However, when the code to vectorize is irregular or with complex dependencies between the variables, the compilers tend to fail. 
Nevertheless, this strategy allows having a single source code clear and portable and still obtain good performance thanks to the compiler. 
This paper presents a new strategy to vectorize irregular and unstructured computing kernels. 
Our method aims to complement existing mechanisms by focusing on non-contiguous data access patterns. 
It allows non-expert programmers to benefit from vectorization and experts to delegate complex vectorization implementations that require many data transformations. 
Indeed, our tool attempts to minimize the number of instructions at the cost of complex vector transformations (permutations, extractions, merging, etc.) that are usually tedious to manage with intrinsics or in assembly.

The contribution of our work is the following:
\begin{itemize}
\item to provide a new automatic vectorization strategy. Our approach operates on an instruction graph, which is transformed to minimize the total number of instructions;
\item to decompose the problem into several sub-problems, and provide different heuristics to solve them;
\item to study the performance of the resulting vectorized kernels against automatic vectorization in modern compilers, namely GCC and icpx Intel, and state-of-the-art techniques in LLVM's vectorizer;
\item to propose a new instruction ordering strategy to reduce the number of register spilling, i.e. the transfers between the stack and the registers.
\end{itemize}

The paper is organized as follows.
In Section~\ref{sec:problem} we detail the background on vectorization and automatic vectorization.
Then, we discuss the related work in Section~\ref{sec:relatedwork}.
We present our method in Section~\ref{sec:Autovesk}.
Finally, we evaluate the benefit of our method in Section~\ref{sec:perfstudy}.

\section{Problem description}
\label{sec:problem}

\subsection{Vectorization}

The principle of vectorization, as developed in modern CPUs, can be described as a technique that allows applying a single instruction on one or several vectors of native data types.
The implementation of this concept in modern CPUs comes with several constraints.
For instance, the size of the vector is fixed and hardware dependent, even though it is possible to pad the vector with arbitrary values to work on smaller sizes.
The vectors are CPU registers, such that load and store operations must be used to exchange data with the main memory.
The operation set supported by a processing unit to manipulate the vectors is hardware dependent.
More precisely, it is tied to the ISA and each ISA may have operations that are not supported by others.
For example, some can allow shuffling of the data inside the vectors, whereas others might only support shuffling inside half of the vectors.
Some might support non-contiguous memory accesses with gather/scatter operations, but others might not.
However, we can surely consider that they all support arithmetic/logical operations, which are usually applied term by term on two vector operands.
In our approach, we consider that vector transformation instructions are available to shuffle the data, extract elements, and merge two vectors in one.
When considering performance, it is usually more efficient to loads/stores contiguous elements in memory and use aligned memory accesses (i.e. the starting address pointer is a multiple of a given value).
Predication is a feature supported by some ISAs, such as ARM SVE, which allows operations to be applied only to certain active elements using predicate vectors. Other ISAs, namely x86, can achieve similar functionality but with a different implementation. It generates mask vectors that are used with binary operations to eliminate inactive elements.

\subsection{Automatic vectorization}

Automatic vectorization consists of converting a scalar source code or binary into a vectorized equivalent without requiring explicit guidance from the developers.
This is a difficult problem because the original program can be written in a way that makes it difficult to vectorize.
The main reasons are the use of data structures/accesses that are not adequate, or when it is difficult to predict the dependencies leading to the impossibility to generate a vectorized kernel before the execution.
Moreover, even if vectorization can be applied, the result can be inefficient and take more time than the original version to execute.
Indeed, if we suppose that we put a single scalar value (or a few of them) inside each vector, it seems straightforward to obtain a naive vectorized program.
However, the waste in the vectors and the fact that vectorized operations can have a higher cost (either in latency or throughput) would lead to poor implementation.
This is why a good metric to predict the performance of a vectorized kernel is to count the number of instructions, with potentially a cost associated with each of them, and consider that the lower is the better.
A more refined optimization can also predict the optimal use of hardware units by considering that some operations applied to disjoint data can be performed concurrently.

In the current study, we focus on static kernels, which allows us to know the original number of scalar instructions.
Also, some non-static kernels could be seen as a dynamic repetition of a static kernel, such that our method is also useful in this case.
To illustrate the difficulty of the problem, we will provide a rough complexity to obtain an optimal solution if we attempt to compute all possibilities to find the optimum.

Consider we have a direct acyclic graph $G(V, E)$ where the $V$ vertices correspond to the $N$ scalar operations of the original program, and the $E$ edges are the dependencies.
We state that the number of possibilities is bounded by
\begin{equation}
    \prod_{i=0}^{\lceil N/k \rceil } \frac{(N - i \times k)!}{(N - k \times i  - k)!} \,
\end{equation}
which is equal to the number of possibilities to group the elements in subsets of $k$ among $N$.
We can lower this number by considering that $N$ is composed of $L$ loads, $S$ stores and $O$ different operations each of $I$ scalar operations (the operations of the same type that are at different depth in the graph can be considered of different type).
However, this limit considers all vectors as full, which might not be the case, increasing, even more, the number of possibilities.
Therefore, a brute force approach to find an optimal vectorial graph seems impossible and motivates the use of heuristics.

\section{Related work}
\label{sec:relatedwork}

Auto-vectorization transforms scalar operations into SIMD operations that can process multiple data elements in parallel, leading to significant performance improvements on modern processors. 
There are two main strategies for automatic vectorization in modern compilers: loop vectorization and Superword Level Parallelism (SLP).

Loop vectorization is a technique that leverages parallelism in loop iterations to transform scalar operations into SIMD operations. 
This process involves expanding each operation in the loop from a scalar type to a vector type, which is a straightforward transformation for simple loops. 
However, many computations require more advanced loop transformations to be in a vectorizable form and preferably without dependencies to ensure consistency. 
One of the challenges of loop vectorization is dealing with non-contiguous access patterns to the operation data. 
To address this, \citet{Nuzman06} introduced an extension to a classical loop-based vectorizer that handles non-unitary computations and data accesses, particularly data with powers of 2 on CPUs. 
This extension helps to exploit spatial locality and address interleaved data access, but remains limited to regular strided access patterns.

Loop vectors are mostly dealing with the data parallelization aspect directly related to the SIMD architecture. 
As a counterpart, there is the SLP algorithm, originally introduced in \cite{Larsen00}. 
It performs instruction-oriented vectorization. 
It is commonly used to convert linear code into vector code, which complements traditional loop-based vectorizers. 
SLP algorithm analyses the input code by scanning the compiler-generated intermediate representation (IR) looking for instruction groups that can be combined into vectors. 
It replaces these selected instruction groups with the corresponding vector instructions. 
\citet{Nuzman07} proposed the Bottom-Up SLP algorithm implemented in the well-known GCC and LLVM compilers.
The main contribution of the algorithm is its ability to handle sequential code anywhere in the program, not just in loops. 
It can also vectorize code inside loops if the loop vectorizer fails. 
Our proposed tool has a similar objective but is currently limited to working with code that uses C++ templates, as it has not yet been integrated into a compiler at the IR level. 
While the evaluation focuses on vectorization opportunities within loops, which are generally more common, it still provides valuable insights into the potential of our techniques.
The SLP approach suffers from some limitations. 
Mainly, it operates locally or even atomically on individual SLP graphs which can be limiting in the case of consecutive graphs sharing data. 
In these cases, an overestimation of the vectorization costs may occur and lead to missed opportunities, even in a simple code.
The SLP vectorizer may miss several opportunities for vectorization in loops. 
\citet{Rodrigo20} presented Vectorization Aware Loop Unrolling (VALU), a new heuristic accompanying the compiler in searching for loop-unrolling opportunities to allow auto-vectorization of sequential code.
VALU checks whether the loop unrolling will be cost-efficient for the SLP vectorizer and identifies the loop unrolling factor that can maximize the use of the vector units in the target architecture.
In Autovesk, we fully unroll the kernels as we create the scalar instruction graph, thus allowing more instruction grouping possibilities, i.e. better vectorization of the code.
\citet{Vasileios18} presented Look-Ahead SLP (LSLP), an improved vectorization algorithm based on SLP that mainly focuses on commutative operations. 
It is implemented in LLVM compiler. 
LSLP's main contributions are first, to be able to extend SLP graph data structure to form multi-nodes of chains of commutative operations of the same type.
Secondly, it introduces a more powerful operand reordering scheme that makes informed decisions based on instructions deeper in the code. 
Our method enables chain reduction for improved vectorization efficiency by transforming the scalar instruction graph and grouping and annotating commutative operation nodes.
\citet{Porpodas19} go further and present Super-Node SLP (SN-SLP), a new algorithm similar to LSLP with multi-nodes, including both commutative operations and their corresponding inverse elements, for example, addition and its inverse subtraction.
SN-SLP leverages the algebraic properties of these operators to allow additional transformations, increasing the possibilities of vectorization.
It is implemented in LLVM. 
Several works have proposed heuristics to optimize instruction ordering and code regions for vectorization in SLP-based algorithms. 
Nevertheless, our new approach is competitive and offers additional flexibility by allowing the vectorization of kernels with irregular data access patterns through instruction grouping and reordering transformations and heuristics.

As vectorization can be seen as a parallelization at the instruction level, the work on automatic parallelization using the polyhedral model can be used~\cite{5260526,10.1007/978-3-540-24644-2_14,inproceedingsBastoul}.
Polyhedral compilers are designed specifically to support calculations in the affine domain in which the loop boundaries and the subscript expressions can be expressed as integer linear functions of both the loop indices and the loop constants. 

Irregular applications such as graph algorithms, particle simulation codes and sparse matrix codes that use compact representations~\cite{10.1145/2925426.2926285, 10.1145/2908080.2908111, 10046127, 10.1145/996841.996853} are now being considered for acceleration using SIMD capabilities. 
Although, the initial focus of SIMD devices was on speeding up regular applications such as dense matrix multiplication and image processing.
Vectorization becomes particularly challenging when dealing with irregular access patterns. 
Another challenge is dealing with memory alignment, which has been largely ignored in previous research focusing on properly aligned memory references. 
Existing vectorization approaches for such irregular applications include inspector/executor and conflict masking. 
Inspector-executor transformations involve loop and data layout transformations that require a runtime component due to unresolved indirect array accesses.
The Sparse Polyhedral Framework (SPF) extends the capabilities of the Polyhedral Framework by incorporating uninterpreted functions to represent non-affine array accesses, non-affine loop boundaries, and inspector-executor transformations~\cite{spf18, spf22}. 
Conflict masking is an approach to resolving data conflicts in SIMD vectors by identifying non-conflicting lanes and writing only to them in each round of execution. 
A conflict detection instruction (vpconflict) has been added to the Intel AVX-512 instruction set to facilitate conflict masking.
The vectorization performance of conflict-masking is dependent on the input distribution. 
If conflicts arise frequently in the input, conflict-masking will result in low SIMD utilization and thus poor vectorization performance. 
\citet{10.1145/3168827} propose a code transformation called in-vector reduction that can efficiently vectorize a class of associative irregular applications. 
Further research aims to support aggressive SIMD vectorization of important loops, even when their bodies contain complex control flow and the entire computation cannot be fully parallelized such as speculative execution~\cite{10.1145/2838734, 6618831}.
Beyond this, \citet{10.1145/3445814.3446692} propose VeGen, an extendable framework that targets non-SIMD vector instructions by introducing a new model of vector concurrency called lane-level concurrency (LLP). 
This new model goes beyond traditional SLP by allowing instructions to perform multiple non-isomorphic operations, and operations on any output lane can use values from arbitrary input lanes. 

\section{Autovesk}
\label{sec:Autovesk}

\subsection{Transformation flow}

\begin{figure}[h!t]
    \centering
    \includegraphics[width=\textwidth]{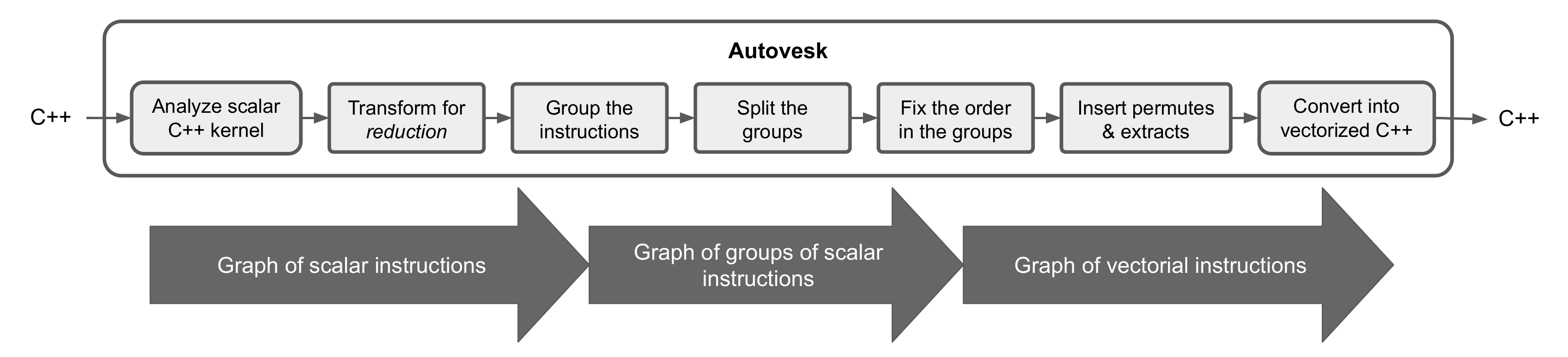}
    \caption{Overview of the transformation flow in Autovesk.
             Starting from a C++ code, the tool uses successively a graph of scalar instructions, a graph of groups, and a graph of vectorial instructions.}
    \label{fig:autoveskflow}
\end{figure}

We provide in Figure~\ref{fig:autoveskflow} an overview of the internal organization of our engine\footnote{\url{https://gitlab.inria.fr/bramas/autovesk}}. 
First, the process starts by obtaining a graph of scalar instructions of a given computational core (Section~\ref{sec:scalar}).
In the second step, the graph can be transformed to exploit reduction when relevant (Section~\ref{sec:reduction}).
In the third step, the instructions are grouped to obtain a meta-graph (Section~\ref{sec:group}).
Next, the groups are split to match the desired vector size denoted by $VEC\_SIZE$ (Section~\ref{sec:size}).
We first divide the loads/stores, then the arithmetic/logical operations. Afterward, the order of the elements in each group is set, and the necessary permutations/extractions are performed (Section~\ref{sec:order}).
Finally, the resulting vector operation graph is ready to be translated into a vector instruction list by our backend (Section~\ref{sec:back}).

To illustrate these different mechanisms, we will consider the vectorization of two kernels given in Figure~\ref{code:examples} under the names $KA$ and $KB$.

\begin{figure}[H]
\centering
\begin{lstlisting}[escapechar=|]
void Kernel_A(const double* inValue1, const double* inValue2, double* outValue, const long size){
    for(long int idx = 0 ; idx < size ; ++idx){
        outValue[idx] = inValue1[(idx+2)%size] * inValue2[(idx*2)%size]; |\label{code:examples:loadstore1}|
    }
}
void Kernel_B(const double* inValue1, const double* inValue2, double* outValue, const long size){
    double x = 0; |\label{code:examples:set}|
    for(long int idx = 0 ; idx < size ; ++idx){
        x += inValue1[idx] + inValue2[idx]; |\label{code:examples:load2}|
    }
    outValue[0] = x;|\label{code:examples:store2}|
}
\end{lstlisting}
\caption{Two example functions to be vectorized which we denote by $KA$ and $KB$.
         A static version of these functions is obtained by fixing the $size$ variable.
         In this case, the loops can be fully unrolled and vanish.}
\label{code:examples}
\end{figure}

\subsection{Scalar graph generation}
\label{sec:scalar}

In our current engine, we generate a direct acyclic graph of scalar instructions using a custom C++ tool based on templates and operator overloading, which builds the graph during the execution of a compiled program. 
This stage will disappear when Autovesk is ported into an existing compiler, but it allows us to validate the transformation steps.
We have four types of scalar nodes that compose our graph:
\begin{itemize}
    \item set: is the assignment of a variable with a given value, for example, Line $\ref{code:examples:set}$ in Figure~\ref{code:examples}.
    \item load: is the access to memory using an address to read a value, for example, Lines $\ref{code:examples:loadstore1}$ and $\ref{code:examples:load2}$ in Figure~\ref{code:examples} (right hand side).
    \item operation: is the use of one or several elements as input and the creation of an output.
    The arithmetic operations belong to this category.
    \item store: is the access to memory using an address to write a value, for example, Lines $\ref{code:examples:loadstore1}$ and $\ref{code:examples:store2}$ in Figure~\ref{code:examples} (left hand side).
\end{itemize}

\begin{figure}[!ht]
    \centering
    \begin{subfigure}[c]{\textwidth}
    \includegraphics[width=\textwidth]{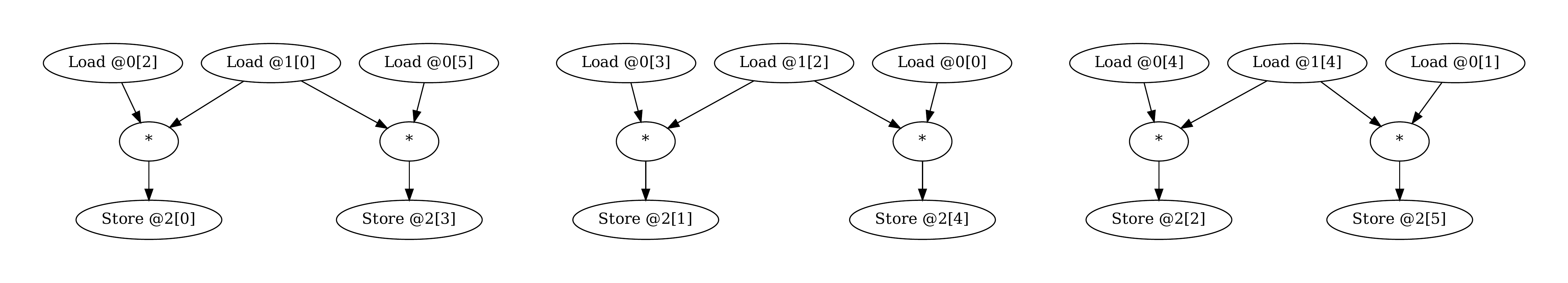}
    \caption{Scalar graph for $KA$}
    \label{fig:scalargraph:KA}
	\end{subfigure}
	
	\begin{subfigure}[c]{\textwidth}
    \includegraphics[width=\textwidth, height=.35\textheight, keepaspectratio]{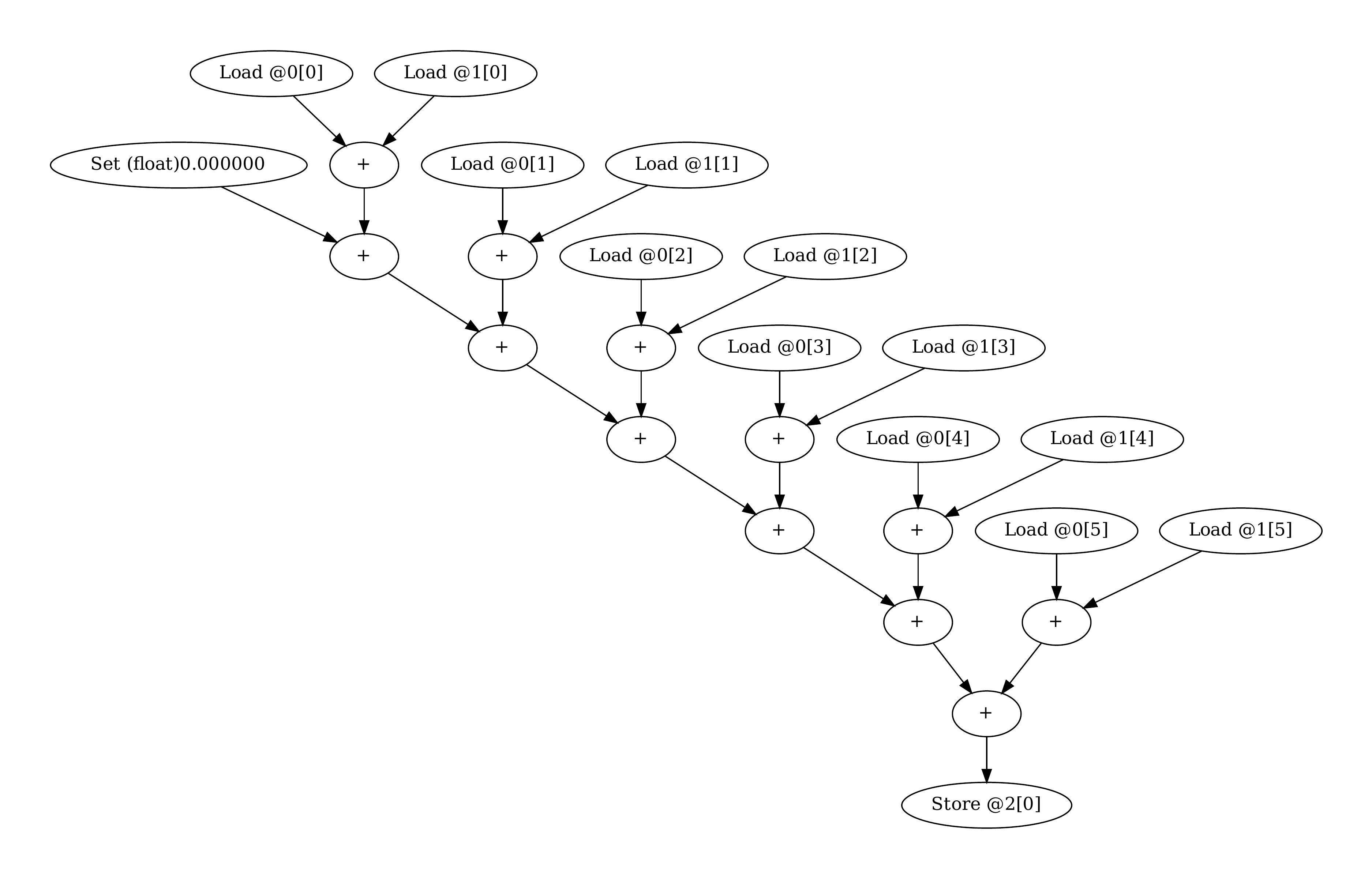}
    \caption{Scalar graph for $KB$}
    \label{fig:scalargraph:KB}
	\end{subfigure}
    \caption{Graphs of scalar instructions for $KA$ and $KB$ for kernel size equals 6.
    The arrays given in the parameters are numbered from 0 to 2 in order.}
    \label{fig:scalargraph}
\end{figure}

What we get is a graph and not a tree since for a given node, the sub-graphs of its successors are not necessarily disjoint.  
For instance, if we have the instructions $a=x$, $b=a*a$, $c=a+a$ and $d=b+c$, the nodes that represent $b$ and $c$ will have the same predecessors and successors.
In our representation, the scalar graph describes operations that are applied from loads/sets to stores without having the notion of variables. 
Thus, some instructions are invisible, such as copying a variable to a local variable, because they do not affect memory.
The graph for $KA$ and $KB$ are given in Figure~\ref{fig:scalargraph}.

At this stage, we can already remove the duplicate nodes.
Two nodes are considered equivalent if they perform the same operation and have the same input.
If the operation is commutative, we consider that the input order is irrelevant.
Therefore, we iterate on the graph from loads/sets to stores and remove the duplicate node wherever possible.

\subsection{Commutative operations and reductions}
\label{sec:reduction}

\begin{figure}[h!t]
    \centering
    \includegraphics[width=\textwidth, height=.35\textheight, keepaspectratio]{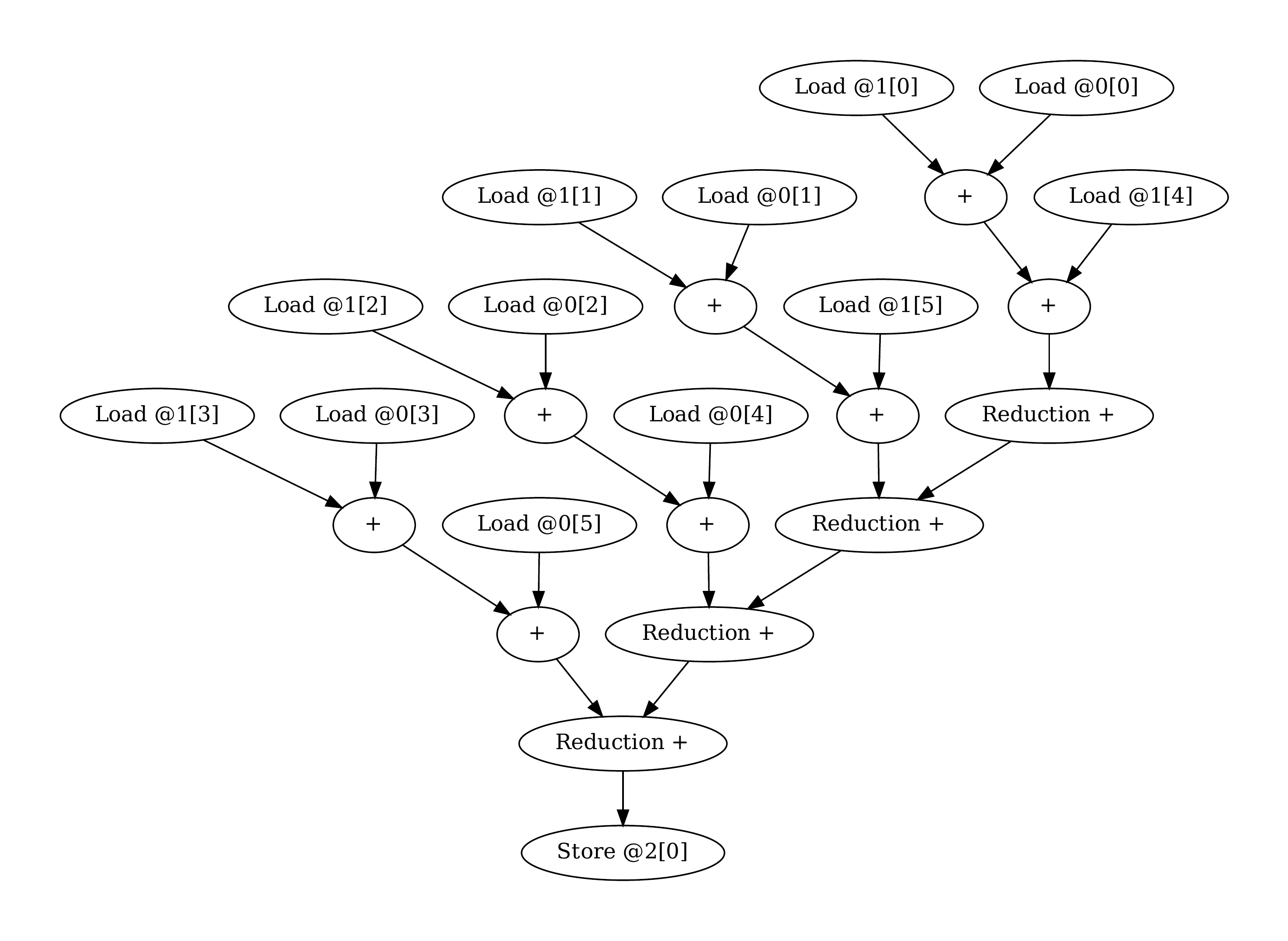}
    \caption{Graph of scalar instructions for $KB$ for kernel size equals 6 and $VEC\_SIZE=4$.
    We can observe that the operations just before the store node were replaced by reduction nodes which materialize the reduction instruction that will be generated later. 
    We can also see that the sources of these nodes are 4 reduction chains (as much as $VEC\_SIZE$) that happen to be isomorphic so they can be vectorized.}
    \label{fig:scalargraph:KB:reduc}
\end{figure}

In vectorization, the reduction pattern consists in changing the order of commutative operations to maintain multiple intermediate results that are then reduced into one, opening the possibility to vectorize more efficiently.
In our case, we enable the reduction by transforming the scalar instruction graph and annotating the corresponding nodes to group them (Section~\ref{sec:group}).
The transformation consists in finding a path in the graph of similar consecutive commutative operations and splitting this path such that we obtain independent $VEC\_SIZE$ sub-paths.
Therefore, we iterate over all the nodes of the graph and when we find a commutative operation, we evaluate if it is the starting point of a reduction path.
Then, we check if the length of the path is enough to hope to benefit from the reduction, i.e. if there are more than $VEC\_SIZE$ nodes.
If it is the case, we split this path into chunks of $VEC\_SIZE$ and connect them with reduction nodes.
We provide the graph for $KB$ after the reduction transformation in Figure~\ref{fig:scalargraph:KB:reduc}. 
Unlike $KB$, there is no opportunity for reduction in the $KA$ example, i.e. there is no chain of commutative operations that are reduced in an output result.

\subsection{Instruction grouping}
\label{sec:group}
\begin{figure}[ht!]
    \centering
    \begin{subfigure}[b]{.48\textwidth}
    \includegraphics[width=\textwidth]{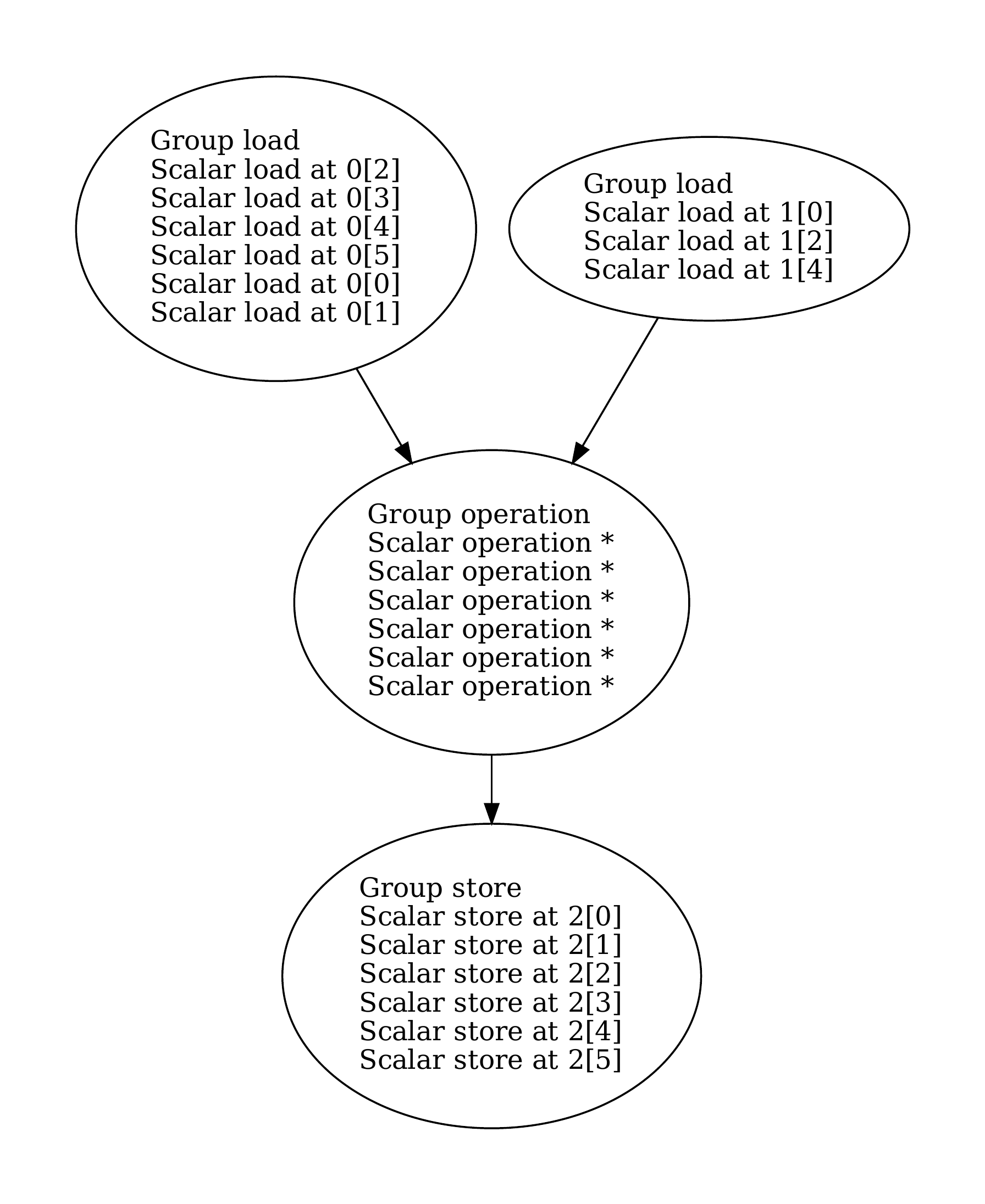}
    \caption{$KA$}
    \label{fig:groupgraph:KA}
	\end{subfigure}
	\hfill
	\begin{subfigure}[b]{.48\textwidth}
    \includegraphics[width=\textwidth, keepaspectratio]{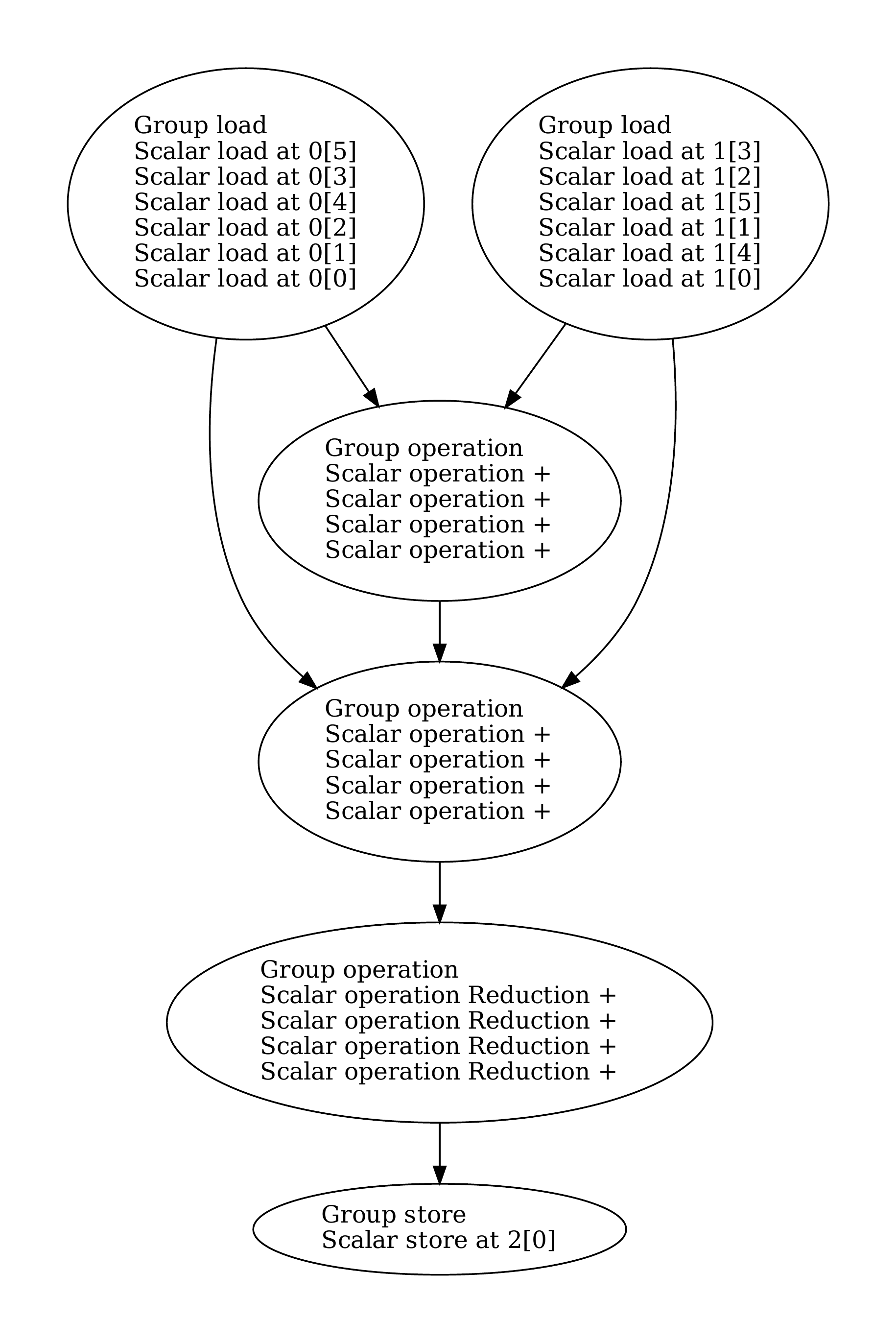}
    \caption{$KB$}
    \label{fig:groupgraph:KB}
	\end{subfigure}
    \caption{Graphs of groups for $KA$ and $KB$ for a kernel size equal to 6.
    An edge between two nodes indicates that at least one instruction from each of the two groups are connected.}
    \label{fig:groupgraph}
\end{figure}

In this step, we generate a meta-graph where the nodes contain lists of scalar operations that could potentially be vectorized together. 
To perform this transformation, we group nodes of the same type.
Thus, we group the loads related to the same array, and we proceed in the same way with the stores. 
Then, we put together the same operations that are at the same distance from the leaves of the graph (i.e., loads and stores). 
By proceeding this way, we ensure that there are no dependencies between the nodes of a group because we are managing it by construction. 
At this stage, groups can contain more scalar nodes than the size of the SIMD vector can handle ($VEC\_SIZE$), and the internal order of scalar operations in the list is unspecified. 
We provide the group for $KA$ and $KB$ in Figure~\ref{fig:groupgraph}.

\subsection{Group divisions}
\label{sec:size}
Once we have the meta-graph, we need to decide how to split the groups that contain more than $VEC\_SIZE$ instructions. We do this in two steps, first on loads/stores, then on operations.

\paragraph{Loads/stores divisions}

To divide the groups of load operations, we follow three rules. 
First, we need to load the values in a contiguous order. 
Second, we want to minimize the number of memory accesses. 
Finally, there should be no more than one vector that is not fully filled, i.e. all used vectors are filled with VEC\_SIZE values and at most, one vector has less than VEC\_SIZE values.
Suppose we have an array of L scalar loads, where $L = k \times VEC\_SIZE + r$, with $0 \leq r < VEC\_SIZE$.
If $L$ is divisible by $VEC\_SIZE$ without leaving a remainder (i.e. $r=0$), there is no choice: loads are in order and all the vectors will be full.
However, when we have $r > 0$, we need to decide which vector will not be fully filled, i.e. how to distribute the scalar loads across the vectors with respect to the contiguous order.
This decision affects the other stages of vectorization because it determines the initial state of the vectors, which in turn affects the number of transformations required in subsequent stages. 
This can have an impact on the final quality of the vector implementation.

\begin{figure}[h!t]
    \centering
    \begin{subfigure}[c]{.48\textwidth}
    \includegraphics[width=\textwidth, page=1, trim=0cm 8cm 5cm 0cm, clip]{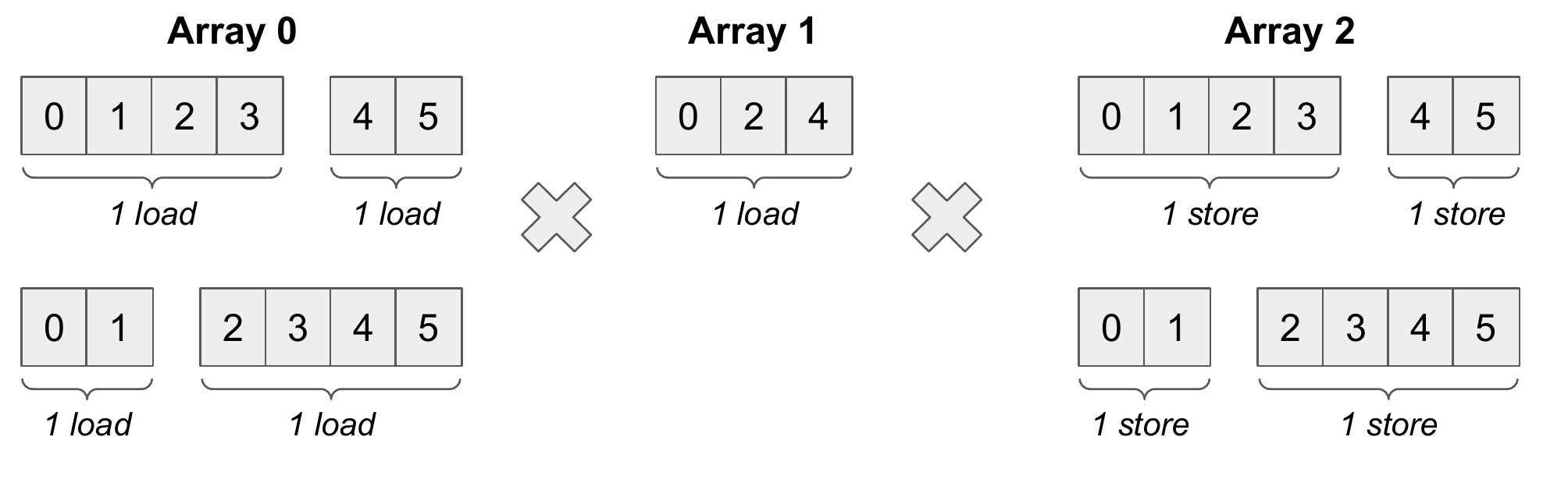}
    \caption{$KA$}
    \label{fig:configs:KA}
	\end{subfigure}
	\hfill
	\begin{subfigure}[c]{.48\textwidth}
    \includegraphics[width=\textwidth, keepaspectratio, page=2, trim=0cm 8cm 5cm 0cm, clip]{./autovesk-possibilites.pdf}
    \caption{$KB$}
    \label{fig:configs:KB}
	\end{subfigure}
    \caption{All the combinations of loads/stores for the two kernels $KA$ and $KB$ considering size equals 6 and $VEC\_SIZE=4$.
    There are 4 possibilities for $KA$ ($2 \times 1 \times 2$) and $KB$ too ($2 \times 2 \times 1$).}
    \label{fig:configs}
\end{figure}

Therefore, we tackle this problem by generating all the possible combinations satisfying the three above rules and picking the combination that leads to the vectorial graph with fewer nodes.
If we consider that the number of vectors needed to load $L$ values is $V$ (with $V = k + 1$ when $r \neq 0$, or $V = k$ otherwise), we have $V$ possibilities.
This is because only one vector selected among the $V$ vectors should be incomplete.
If there are $A$ arrays accessed in the kernel, either by load or store operations, the total number of possibilities is given by $C = \prod_{i = 1}^{i = A}{V_i}$, with $V_i$ the number of vectors for array $i$.
Concretely, if we have two arrays as input and one array as output, and load ten vectors from each, $C$ will be equal to one thousand.
The different combinations for $K$ are shown in Figure~\ref{fig:configs}.

\begin{figure}[!ht]
\centering
\includegraphics[width=0.9\textwidth, trim=0cm 5cm 1cm 0cm, clip]{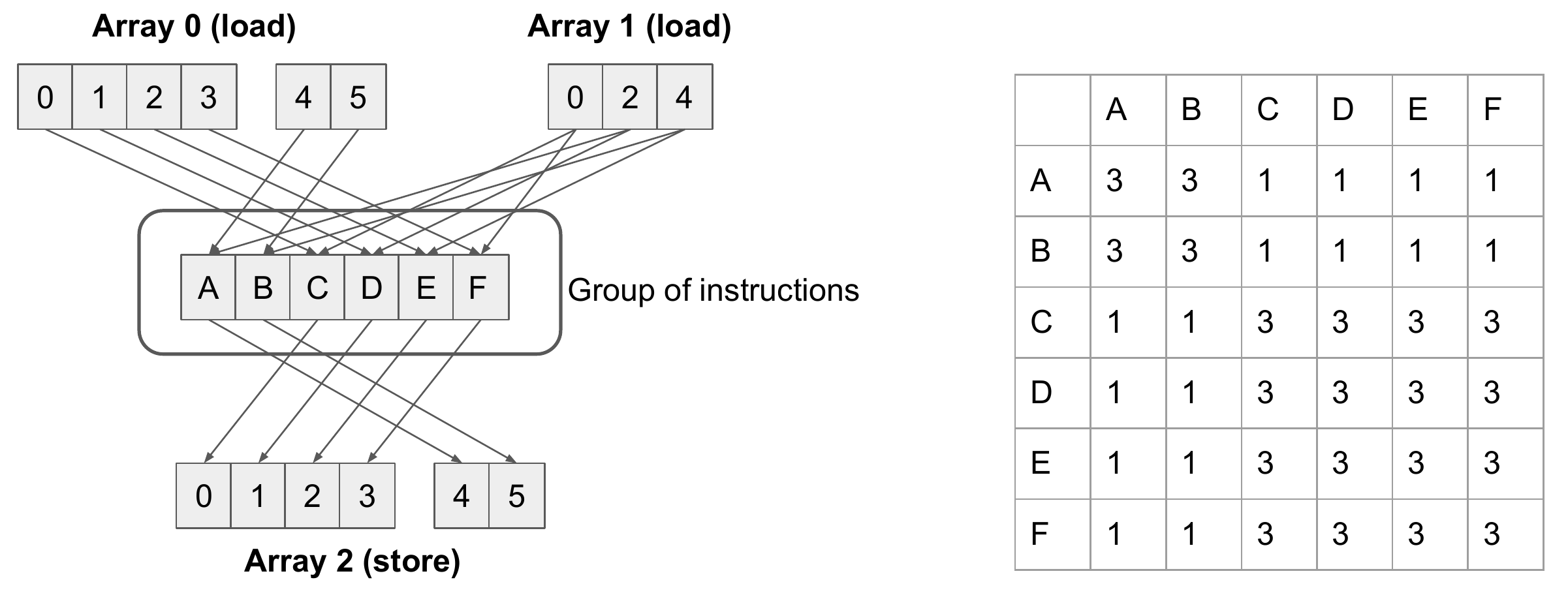}
\caption{Example of score matrix $d$ for $KA$ to split the scalar instruction group.
        The groups of loads and stores are already divided, and the Algorithm~\ref{algo:score-a-b} is used to compute the matrix.
        The given matrix is found out for this specific configuration of loads/stores and could be different otherwise.}
\label{fig:exemple-score}
\end{figure}

\paragraph{Operation divisions}
After splitting the load/store groups, we split the operation groups. We iterate over the operation groups from loads to stores. By going through the operation groups this way, we can be sure that the inputs to the operations are already divided (or if the inputs are loads, they are also already fixed).
For each group $g$ we compute a score-matrix $d$ of dimension $|g| \times |g|$ that aims to express which of the elements of $g$ should be in the same vectorial operation, i.e. $d(i,j)$ describes the expected interest of putting the scalar operations $i$ and $j$ in the same vectorial operation.
The calculation of this score is defined in Algorithm~\ref{algo:score-a-b}.
The score $d(i,j)$ increases with the number of common predecessors/successors between $i$ and $j$.
If they are two separate nodes, we increment the score by the number of sources in common.
For each destination group: if it is a \emph{store} group, we increment the score by 1; if it is an \emph{operation}, we increment by 1 in the case of a group of size less than or equal to $VEC\_SIZE$, or we increment by the inverse of the size of the destination in the case of a group of size greater than $VEC\_SIZE$, so the larger the size of the destination, the less value we bring to this group.
We show an example of a score-matrix in \autoref{fig:exemple-score}.

\begin{algorithm}[!ht]
 \caption{Compute the matrix entry $d(A,B)$ for elements $A$ and $B$ of the same group.}
 \label{algo:score-a-b}
    $score \gets$ 0\;
    \If{A $\neq$ B}{  \tcp{Increment score by 1 or 2 for same sources} 
        score += (in\_same\_vector(A.src1,B.src1) + in\_same\_vector(A.src2,B.src2))\;
    }
    \Else{
    \tcp{For the diagonal, add the number of sources, i.e. 1 or 2}
    score += (A.src1 != null) + (A.src2 != null)\;
    }
    \tcp{Increment score if they are used at the same place (destination)}
    \For{d \textbf{in} Intersection(A.dest,B.dest)}{
        \uIf{d is a Store Group}{
         score += 1\;
        }
        \Else{
            \tcp{d is an Operation Group}
            \If{d.size $\leq VEC\_SIZE$ }{
            score += 1\;
            }
            \Else{
            score += (1 /d.size)\;
            }
        }
    }
\end{algorithm}

In order to split the group of operations, we propose two main strategies based on dividing the matrix $d$:
\begin{enumerate}
    \item a clustering strategy where we create initial sub-groups and then aggregate the elements using the best score remaining so far;
    \item a partitioning strategy where we split the matrix at the point with the lowest score.
\end{enumerate}
We know we will need $V = \left \lceil |g|/VEC\_SIZE \right \rceil$ vectors.
The partitioning strategy is provided in Algorithm~\ref{algo:partitioning} in Appendix~\ref{app:partitioning}.
We start with a partition that contains all the elements.
We find in this partition the two elements with the lowest score in the matrix (line~\ref{algo:partitioning:s3}) and initiate two new partitions (line~\ref{algo:partitioning:s4}).
Then we aggregate the remaining elements on these two partitions while ensuring that no partition exceeds a given size. 
This is done either by finding the highest accumulated score among the remaining elements to move them one by one (line~\ref{algo:partitioning:s5}) or by placing each element in the partition where it has the heaviest weight overall, i.e. the maximum accumulated sum of the scores of an element relative to all the others.

The clustering strategy is provided in Algorithm~\ref{algo:clustering} in Appendix~\ref{app:clustering}.
We start by finding the two elements $i$ and $j$ with the lowest scores in the matrix, i.e. those with the least number of predecessors and successors in common, and use them to initiate two sub-groups (line~\ref{algo:clustering:s1}). 
Then, we find $V-2$ other elements, one by one, by selecting the elements having the lowest scores compared with the already fixed elements (line~\ref{algo:clustering:s2}).
We end up with $V$ sub-groups that each contain one element.
We finish by processing the remaining elements. We compute, for each of them, the sum of the scores of the nodes of all subgroups respectively and choose the sub-group having the best score (line~\ref{algo:clustering:s3}).

\subsection{Fixing the order of vector's elements}
\label{sec:order}

At this stage, we have sub-groups that contain at most $VEC\_SIZE$ scalar instructions. 
The order of these instructions is fixed for the loads and stores but not for the operations. 
That is the aim of this layer. 
We have to find the correct positions to make the operations coincide with their predecessors and successors; thus, we will ensure the consistency of the kernel. 
Fixing the order will imply using permutations, merges and extractions; therefore, the objective is to find each group order to minimize the need for additional instructions. 
We note that using an \textit{Extract} instruction is costly in terms of execution time for the vectorized code.
We also note that if we have to extract a single element through a vectorial instruction, this is equivalent in terms of execution time to extracting several elements less than or equal to the size of the vector instruction.

\begin{figure}[!ht]
    \centering
    \begin{subfigure}[c]{.48\textwidth}
    \includegraphics[width=\textwidth,page=1, trim=0cm 9.3cm 7.5cm 0cm, clip]{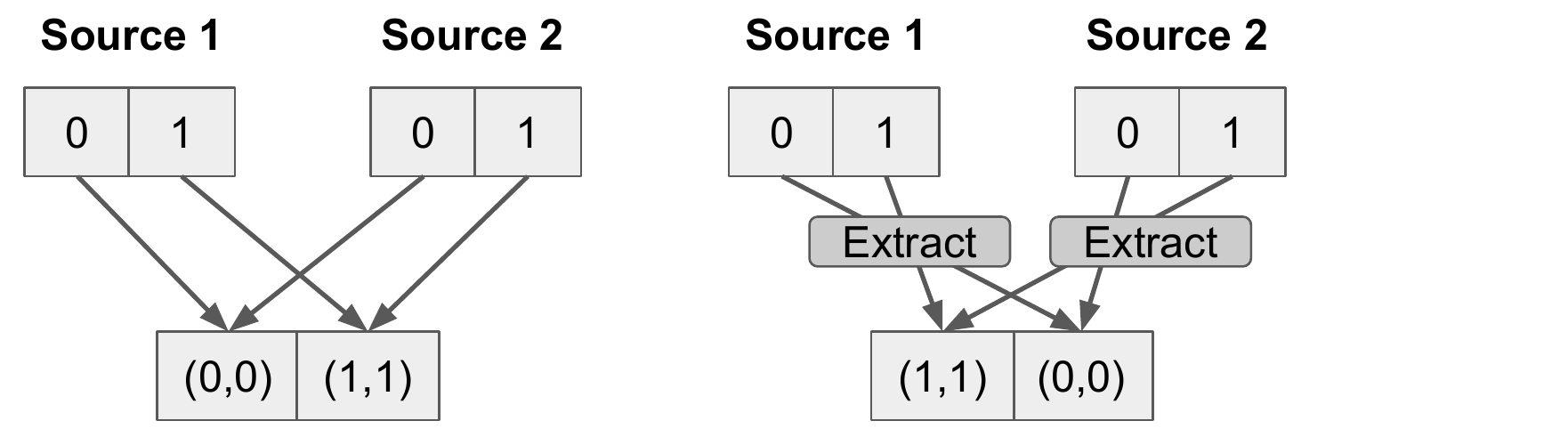}
    \caption{}
    \label{fig:order:1}
	\end{subfigure}
	\hfill
	\begin{subfigure}[c]{.48\textwidth}
    \includegraphics[width=\textwidth,page=2, trim=0cm 9.3cm 7.5cm 0cm, clip]{./autovesk-order}
    \caption{}
    \label{fig:order:2}
	\end{subfigure}
	
    \begin{subfigure}[c]{.58\textwidth}
    \includegraphics[width=\textwidth,page=3, trim=0cm 7.5cm 0cm 0cm, clip]{./autovesk-order}
    \caption{}
    \label{fig:order:3}
	\end{subfigure}
    \caption{Three examples with different order of the elements of the output vector.
    An \textit{Extract} operation is used to select some elements and/or permuting elements.
    A merge operation is used to select elements from two vectors and store them into a single one.}
    \label{fig:order}
\end{figure}

In a sub-group, for $VEC\_SIZE$ scalar operations, we have $VEC\_SIZE!$ possibilities to order them, leading to the impossibility to compare all of them greedily.
That said, in our case, an additional layer of constraints is added, which reduces the number of such possibilities.
We have the connections between a sub-group of operations and its predecessors and successors, the order of the predecessors is already fixed (the order of a successor is fixed if it is a store).
Therefore, taking into account the order of a successor, for example, leads to being forced to use extract/permute instructions to retrieve the data required for the operation from other groups and ensure kernel consistency.
To illustrate the impact of a good order of operations, we present the examples in Figure \ref{fig:order}.

It is enough that the positions of the operations in the group of instructions are changed that we find ourselves forced to add instructions \textit{Extract} to have the vectorial instructions coincide. 
A good order prevents this.
To address this issue, we propose Algorithm \ref{algo:score-op-order} which attempts to minimize the number of \textit{Extract} instructions needed for the consistency of a given graph.
If for a given operation node, we have the indices of its two successors equal then we can say that we have the right order and we set the position of this index.
On the other hand, if the indices are different, we no longer have a choice, we set the position of one and admit that the other needs an \textit{Extract}.

\begin{algorithm}[!htbp]
 \caption{Set the order of operations by minimizing the Extracts}
 \label{algo:score-op-order}
    $allNodes \gets$ mapNodesWithIdNode()\;
    \tcp{Add nodes that have connections between each other}
    \For{node \textbf{in} $allNodes$}{
        \For{op \textbf{in} allOperations}{
            $node.depIds \gets$ addDepNode(op)\;
        }
    }
    \tcp{Connect nodes if they are linked by operations (including self-connect)}
    \For{op \textbf{in} allOperations}{
        \For{idx1 \textbf{in} op.depIds}{
            \For{idx2 \textbf{in} op.depIds}{
                \tcp{Not the same index, so fixing one implies extracting to the other}
                \If{idx1 $\neq$ idx2}{
                    ConnectDependantNodes()\;
                }
            }
        }
    }
    \tcp{Sort the nodes by number of constraints}
    sort($allNodes$)\;
    $unusedPositions \gets$ init()\;
    $unusedOperations \gets$ init()\;
    
    \tcp{Fix order of operations that they don't need an Extract}
    \For{node \textbf{in} $allNodes$}{
        $needExtractAnyway \gets$ findIfNeedExtractAnyway()\;
        \If{not needExtractAnyway}{
        fixPosition($unusedPositions$, $node.operation$)\;
        $unusedPositions$.erase($node.position$)\;
        $unusedOperations$.erase($node.operation$)\;    
        }
    }
    \tcp{Put operations that will need an Extract anyway in the vacant positions}
    \While{unusedPositions not empty}{
        fixPosition($unusedPositions$, $unusedOperations$)\;
    }
\end{algorithm}

\subsection{Prospecting for the best configuration using a greedy strategy}
\label{sec:greedy}

\begin{figure}[!ht]
    \centering
    \begin{subfigure}[c]{\textwidth}
    \includegraphics[width=\textwidth]{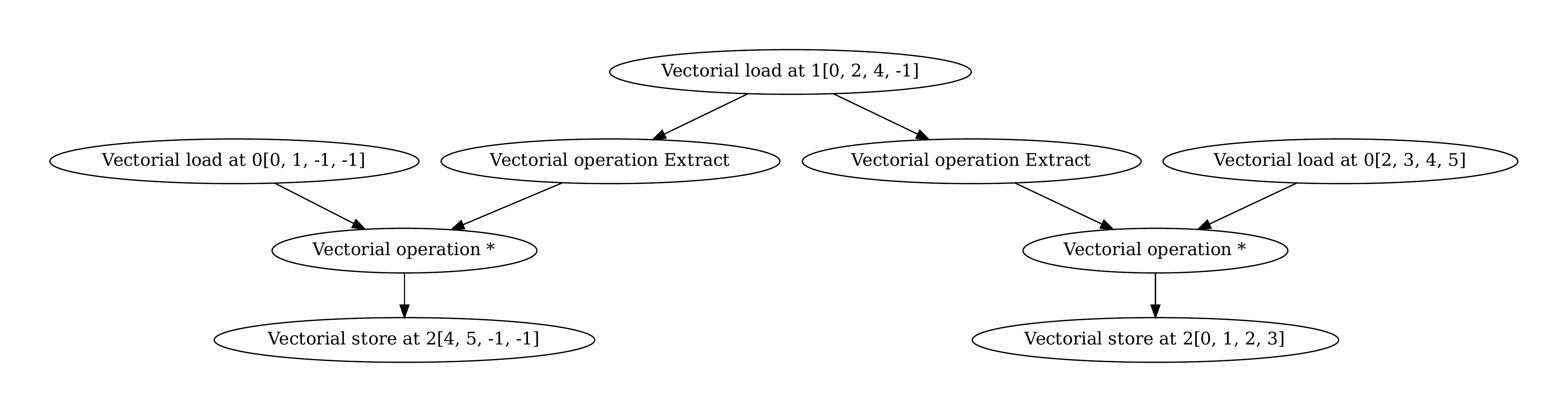}
    \caption{$KA$}
    \label{fig:vecgraph:ka}
	\end{subfigure}

	\begin{subfigure}[c]{\textwidth}
    \includegraphics[width=\textwidth]{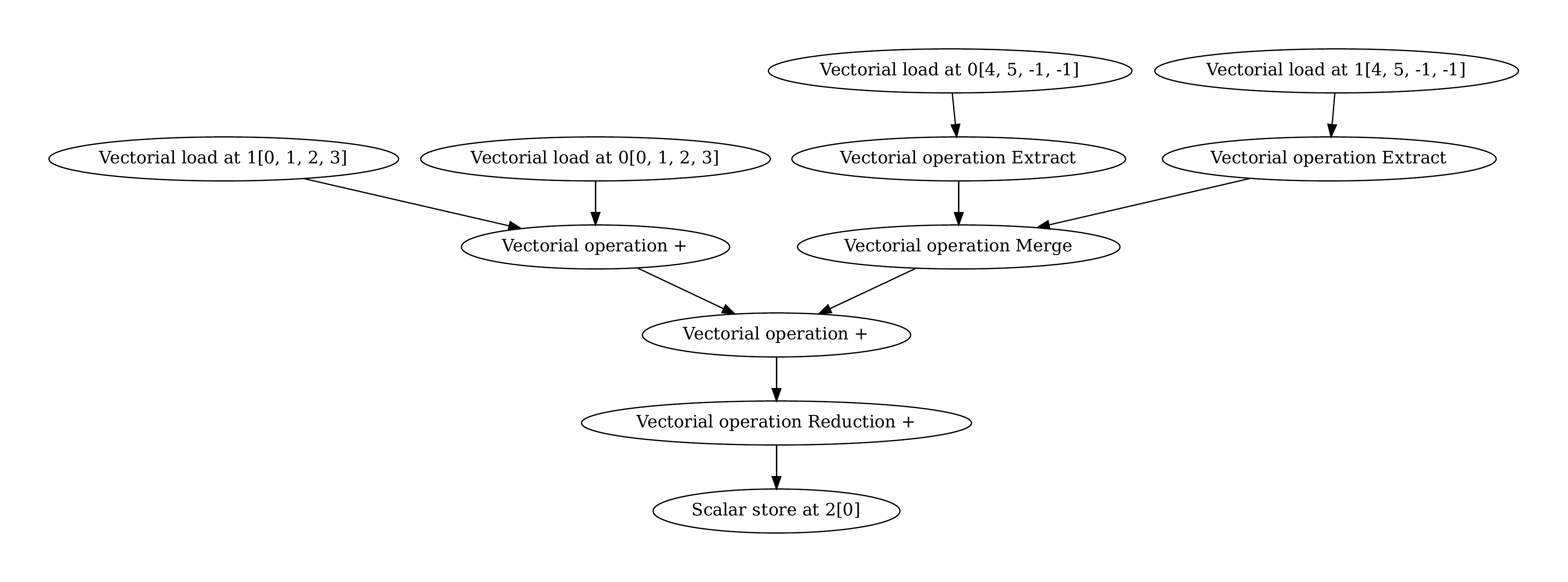}
    \caption{$KB$}
    \label{fig:vecgraph:kb}
	\end{subfigure}
    \caption{Graphs of vectorial instructions for $KA$ and $KB$ for kernel size equals 6 and $VEC\_SIZE=4$.}
    \label{fig:vecgraph}
\end{figure}

In the previous sections, we introduced various strategies for group partitioning, element ordering and optimization. 
However, some of these transformations are exclusive and cannot be used jointly, such as choosing between clustering and partitioning algorithms for operation group splitting. 
Similarly, only one of the split configurations of the loads/stores can be used. 
In addition, not all kernels benefit from reduction transformations, adding to the difficulty of predicting the best approach. 
To meet this challenge, we propose to test all the strategies with a greedy algorithm and select the one that produces the graph with the least number of instructions.
We provide the final vectorial graph we obtain for $KA$ and $KB$ in Figure~\ref{fig:vecgraph}.

In cases where a kernel is complex, exhaustive testing of all transformation configurations may be impractical and time-consuming. 
However, users can guide our tool and explicitly configure the transformations. 
This option allows them to explore the search space selectively and thus save time. 
Determining the optimal configuration depends on the user's understanding of the problem and the specific characteristics of the kernel, such as potential reductions, contiguous load accesses, and the need for advanced group partitioning techniques. 
By considering these factors, users can effectively navigate the transformation options and improve the vectorization process.
In summary, the greedy selection algorithm can be useful to further assist in identifying the most efficient transformation configuration.

\subsection{Backend}
\label{sec:back}

Our backend takes as input the graph of vectorial instructions and generates a C++ program by converting each instruction into an intrinsic function call.
The conversion is straightforward and no low-level optimization is performed at this stage.
However, because we fully unroll and inline the kernels we can potentially obtain a significant code portion which can put pressure on the registers, leading to a register spilling.
Therefore, to help the compiler that will compile the generated C++, we propose a reordering algorithm that aims at reducing the save/restore of intermediate variables.

To this end, we propose the Algorithm~\ref{algo:schedule}. 
It is a two-stage algorithm.
In the first one, we find disjoint parts in the instruction list, i.e. parts of the graph that are not connected.
Each of these parts is then treated separately in the second step.
Our second step consists in iterating over the graph as if it was a graph of tasks.
We maintain a list of ready tasks (the instructions that can be computed), select among them the task that will be computed next and then release the dependencies, with the possibility to move new tasks to the list of ready tasks.
This approach offers several advantages.
It has low complexity and does not require managing the correctness of moving instructions before/after the instructions they depend on (it prevents generating cycles).
In addition, all the core part is then in the selection of the next task among the ones in the ready list.

\begin{algorithm}[H]
 \caption{Schedule a list of instructions $l$.}
 \label{algo:schedule}
    instructions $\gets \emptyset$ \;
    ready\_list $\gets$ roots($l$)\;
    invdepth $\gets$ distance\_from\_leaves($l$)\;
    \While{ready\_list is not empty}{
        costs $\gets$ compute\_costs(ready\_list)\;
        \tcp{Sort the instructions using their costs first and their invdepth if tie}
        ready\_list $\gets$ sort(ready\_list, costs, invdepth)\;
        next\_i $\gets$ ready\_list.front\;
        instructions.insert(next\_i)\;
        ready\_list.pop\_front()\;
        ready\_list $\gets$ manage\_dependencies(next\_i)\;
    }
    return instructions\;
\end{algorithm}

Our strategy consists in sorting the list based on a register benefit and depth to the leaves of the graph.
We present in the Algorithm~\ref{algo:minregisters:cost} the computation of the cost of scheduling an instruction as the next one.
The benefit is based on the number of registers that will be released if the instruction is computed.
If we consider that there are no unused variables, a load or an operation has a cost of 1 because the result of the instruction will have to be stored somewhere.
Then for each instruction, we iterate on its predecessors (input) and see what will be the gain of computing it.
For this purpose, we do the sum of the average of the successors of the predecessors.
We provide an example in Figure \ref{fig:minregisters:example}.
We use the longest distance to the leaves to sort the instructions only when the benefit is equal between several tasks.
Consequently, our algorithm has a local view of the graph and focuses on a minimum local.

\begin{minipage}[t]{.39\textwidth}
    \begin{figure}[H]
    \includegraphics[width=\textwidth,trim=0cm 0cm 11cm 0cm, clip]{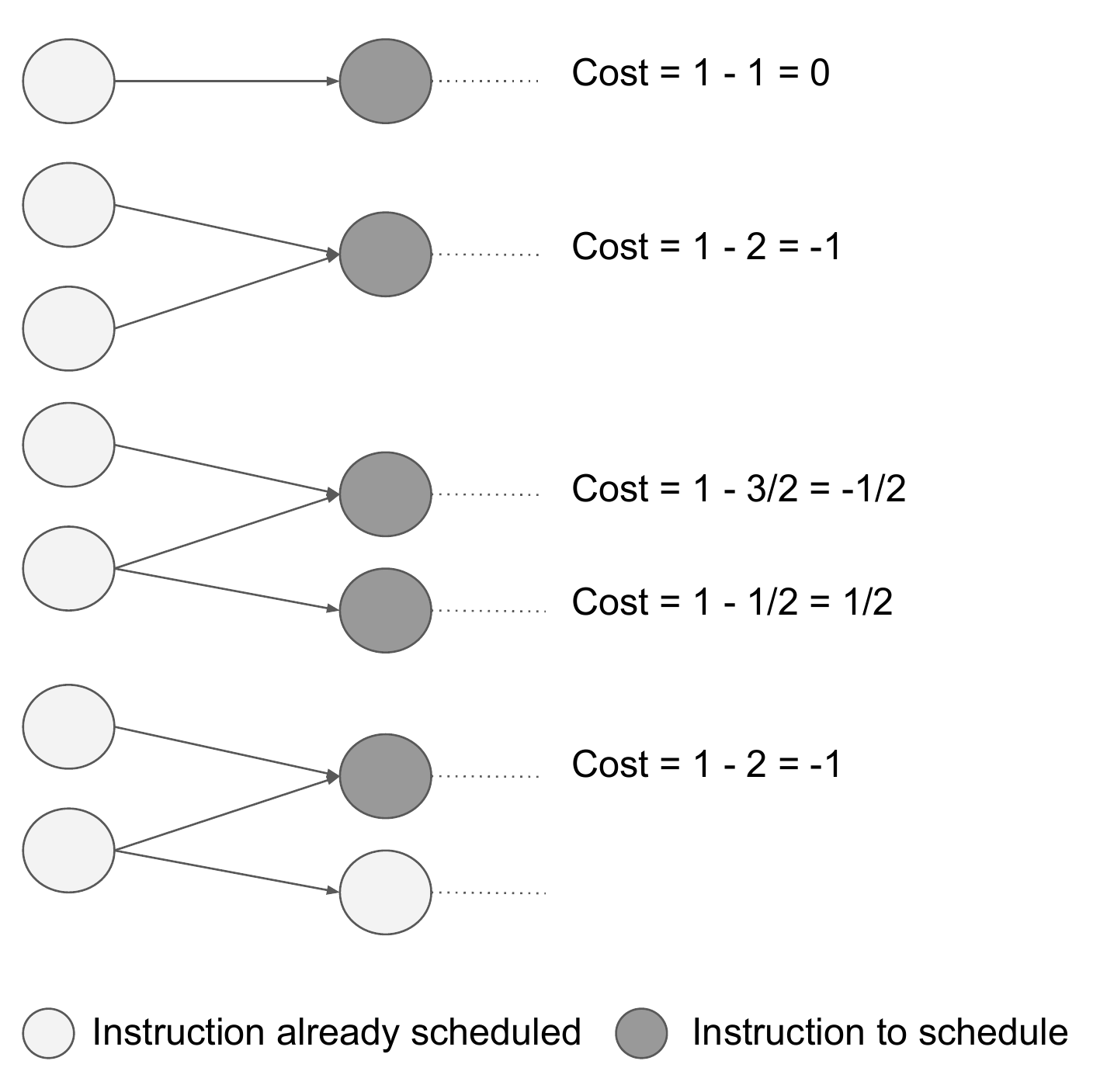}
    \caption{Example of costs to schedule the instructions.}
    \label{fig:minregisters:example}
	\end{figure}
\end{minipage}\hspace{0.2cm}
\begin{minipage}[t]{.57\textwidth}
    \begin{algorithm}[H]
 \caption{Compute the cost of scheduling instruction $i$ next.}
 \label{algo:minregisters:cost}
    cost $\gets$ 0\;
    \If{$i$ has successors}{
    \tcp{We need a register to store the result}
        cost += 1\;
    }
    \For{p in $i.predecessors$}{
        \tcp{Considered $p$ is done already}
        \For{sp in $p.successors$}{
            \If{sp is not done}{
                counter += 1\;
            }
        }
        cost -= 1/counter\;
    }
    return cost\;
\end{algorithm}
\end{minipage}

\section{Performance study}
\label{sec:perfstudy}

\subsection{Configurations}

We evaluate our tool on two platforms:
\begin{itemize}
    \item INTEL-AVX512: is an Intel Xeon Skylake Gold 6240 with 2x 18 cores at 2.6GHz and 512-bit AVX, i.e. a vector can contain eight double floating-point values.
    The node has 192 GB memory and each core has 64KB private L1 cache, 1MB private L2 cache and 1.375MB private L3 cache.
    The OS is CentOS Linux release 7.6.1810 (Core).
    We use the GNU compiler 10.2.0, Clang the LLVM-based compiler 15.0.1 and icpx the Intel compiler 2022.0.0. 
    \item ARM-SVE: is an ARMv8.2 A64FX - Fujitsu with 48 cores at 1.8 GHz and 512-bit SVE, i.e. a vector can contain eight double floating-point values. 
    The node has 32 GB HBM2 memory arranged in four core memory groups (CMGs) with 12 cores and 8GB each, 64KB private L1 cache, and 8MB shared L2 cache per CMG.
    The OS is Red Hat 8.3.1-5.
    We use the GNU compiler 10.3.0 (20210408).
\end{itemize}

We compile both the scalar (original programs) and the vectorized ones (output of Autovesk) with the optimization flags \codebox{-march=native -mtune=native -O3 -ffast-math}.
The executions are pinned to a single core using \emph{taskset}.

\subsection{Test cases}

We use 10 custom computational kernels (Set-CK) and 4 common numerical kernels (Set-NK) to evaluate Autovesk's performance in vectorizing.

Set-CK kernels are specifically designed to cover a wide range of patterns.
They include kernels with scalar input/outputs (denoted $1$) and those with arrays (denoted $N$).
The access patterns to the elements of the arrays vary and can be either contiguous, random or shifted (circular access).
These kernels are in the form of for loops on data arrays. 
They represent a variety of scenarios, including linear contiguous access, irregular access, and circular access.
Each of the 10 kernels in the set consists of two inputs and one output. 
Some of these kernels are expected to be vectorized well by modern compilers.

The Set-CK kernels are:

\begin{tabular}{p{3cm} l}
$\cdot$ (N,N) $\rightarrow$ N: &
    $dest[i] = op(src0[i], src1[i]), \forall i \in [O,N[$. \\
    & This kernel is similar to Blas \emph{axpy}. \\
\end{tabular}

\begin{tabular}{p{3cm} l}
$\cdot$ (N,N) $\rightarrow$ 1: &
    $dest += op(src0[i], src1[i]), \forall i \in [O,N[$. \\
    & This kernel includes KB and is similar to a dot product. \\
\end{tabular}

\begin{tabular}{p{3cm} l}
$\cdot$ (N,1) $\rightarrow$ N: &
    $dest[i] = op(src0[i], src1), \forall i \in [O,N[$ \\
\end{tabular}

\begin{tabular}{p{3cm} l}
$\cdot$ (N,1) $\rightarrow$ 1: &
    $dest += op(src0[i], src1), \forall i \in [O,N[$ \\
\end{tabular}

\begin{tabular}{p{3cm} l}
$\cdot$ (r(N),N) $\rightarrow$ N: &
    $dest[i] = op(src0[r(i)], src1[i]), \forall i \in [O,N[$ \\
\end{tabular}

\begin{tabular}{p{3cm} l}
$\cdot$ (N,N) $\rightarrow$ r(N): &
    $dest[r(i)] += op(src0[i], src1[i]), \forall i \in [O,N[$ \\
\end{tabular}

\begin{tabular}{p{3cm} l}
$\cdot$ (r(N),N) $\rightarrow$ 1: &
    $dest += op(src0[r(i)], src1[i]), \forall i \in [O,N[$ \\
\end{tabular}

\begin{tabular}{p{3cm} l}
$\cdot$ (r(N),1) $\rightarrow$ N: &
    $dest[i] = op(src0[r(i)], src1), \forall i \in [O,N[$ \\
\end{tabular}

\begin{tabular}{p{3cm} l}
$\cdot$ (r(N),1) $\rightarrow$ 1: &
    $dest += op(src0[r(i)], src1), \forall i \in [O,N[$ \\
\end{tabular}

\begin{tabular}{p{3cm} l}
$\cdot$ (s(N),s(N)) $\rightarrow$ N: &
    $dest[i] = op(src0[s(i)], src1[s(i)]), \forall i \in [O,N[$. \\
    & This kernel includes KA. \\
\end{tabular}

The shift function $s(x)$ is defined as
\begin{equation}
    s(x) = (x+2) \pmod N .
\end{equation}
The random access function $r(x)$ is defined as
\begin{equation}
    r(x) = (x\ XOR\ 0x55555555) \pmod N .
\end{equation}
A binary operation $op$ is either a sum or a multiplication.

We evaluate each kernel from size equal to 4 to 128, and an increasing step equal to 4.
An evaluation consists in executing a kernel on 21 distinct arrays 10000 times.
Consequently, in a run, we use at most $21 \times 3$ arrays of 128 double floating point values, which takes $63KB$, such that the data fit in the L1 cache for both hardware configurations.

The kernels from Set-NK come from real applications:
\begin{itemize}
    \item \emph{bcucof} (size 4): \emph{bcucof} routine~\cite{10.5555/1403886} computes a table for bi-cubic interpolation. The bi-cubic algorithm is frequently used for scaling images and video for display.
    \item \emph{bcucofx2f}: It is two consecutive calls to \emph{bcucof} (size 4) on different data that are vectorized together.
    \item \emph{weight} (size 7): \emph{weight} routine~\cite{10.5555/1403886} computes differentiation matrices for pseudo-spectral collocation, which are essential in spectral methods. Spectral methods are a powerful tool widely used for solving partial differential equations (PDEs) due to their high accuracy and efficiency.
    \item \emph{convolution} (sizes 8 and 16) Discrete convolution is a mathematical operation that combines two discrete functions f and g to produce a new function f*g representing their combined effect. It is commonly used for filtering, smoothing and feature detection in various fields including audio and image processing.
\end{itemize}

For all the test cases, the data are initialized before calling the kernels, such that they are already in the cache before the first execution.
When showing the speedup, we obtain it by comparing the scalar version automatically vectorized vs. the version vectorized by Autovesk both compiled with the same compiler.


\subsection{Results}

\paragraph{Set-CK}

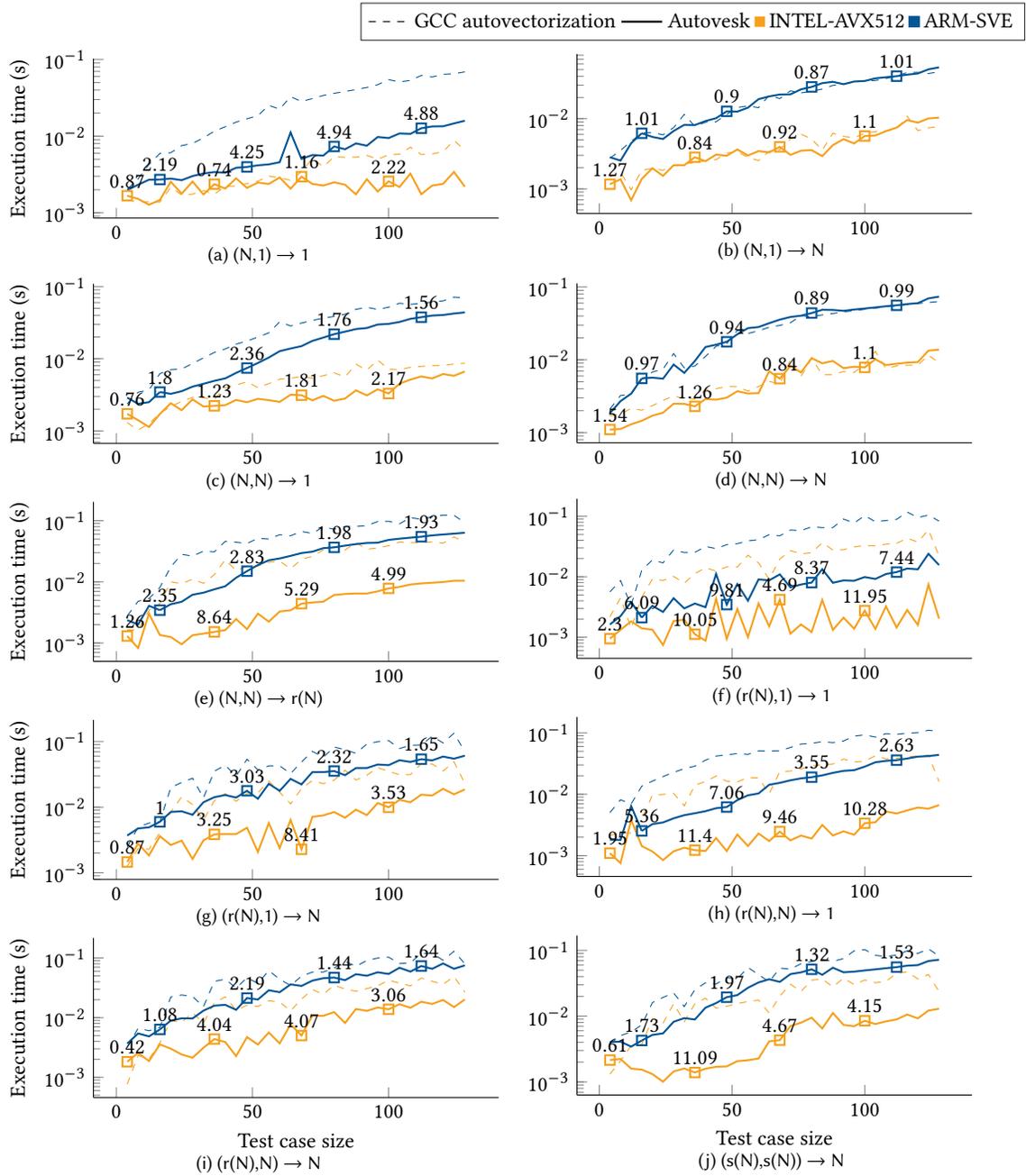
\begin{figure}[h!]
    \centering
    \hfill
        \begin{tikzpicture}
        \begin{axis}[%
                width=.5\textwidth,
        hide axis,
        xmin=10,
        xmax=50,
        ymin=0,
        ymax=0.4,
        legend columns=4,
        legend style={draw=white!15!black,legend cell align=left},
        legend style={at={(0.2,0.)}}
        ]
        \addlegendimage{draw=black, dashed}
        \addlegendentry{GCC autovectorization};
        \addlegendimage{draw=black, thick}
        \addlegendentry{Autovesk};
        \addlegendimage{fill=ACMOrange, draw=ACMOrange, mark=square*, only marks}
        \addlegendentry{INTEL-AVX512};
        \addlegendimage{fill=ACMDarkBlue, draw=ACMDarkBlue, thick, mark=square*, only marks}
        \addlegendentry{ARM-SVE};
        \end{axis}
        \end{tikzpicture}
    
    \begin{subfigure}[c]{0.5\textwidth}
        \begin{tikzpicture}
        	\begin{axis}[
        	    ymode=log,
        	    height=.2\textheight,
                width=\textwidth,
                axis lines*=left,
                ylabel={Execution time (s)},
                 every node near coord/.style={font=\normalsize},
        	]
        	\addplot [draw=ACMOrange, dashed] table [y index= 1,x index=0] {res-avx/N1_1_heap_bench_res_1658625605.txt};

        	\addplot [nodes near coords, draw=ACMOrange, thick, mark=square, mark repeat={8},point meta=explicit] table [y index= 2,meta index=3, x index=0] {res-avx/N1_1_heap_bench_res_1658625605.txt};

        	\addplot [draw=ACMDarkBlue, dashed] table [y index= 1,x index=0] {res-armsve/N1_1_heap_bench_res_1657156478.txt};

        	\addplot [draw=ACMDarkBlue, thick,restrict x to domain=4:16] table [y index= 2,x index=0] {res-armsve/N1_1_heap_bench_res_1657156478.txt};

        	\addplot [nodes near coords, draw=ACMDarkBlue, thick, mark=square, mark repeat={8},point meta=explicit, restrict x to domain=16:128] table [y index= 2,meta index=3, x index=0] {res-armsve/N1_1_heap_bench_res_1657156478.txt};
        	
        	\end{axis}
        \end{tikzpicture}
        \subcaption{(N,1) $\rightarrow$ 1}
        \label{fig:res-n11}
    \end{subfigure}
    \begin{subfigure}[c]{0.49\textwidth}
        \begin{tikzpicture}
        	\begin{axis}[
        	    ymode=log,
        	    height=.2\textheight,
                width=\textwidth,
                axis lines*=left,
                 every node near coord/.style={font=\normalsize},
        	]
        	\addplot [draw=ACMOrange , dashed] table [y index= 1,x index=0] {res-avx/N1_N_heap_bench_res_1658625605.txt};

        	\addplot [nodes near coords, draw=ACMOrange, thick, mark=square, mark repeat={8},point meta=explicit] table [y index= 2,meta index=3, x index=0] {res-avx/N1_N_heap_bench_res_1658625605.txt};

        	\addplot [draw=ACMDarkBlue , dashed] table [y index= 1,x index=0] {res-armsve/N1_N_heap_bench_res_1657156478.txt};

        	\addplot [draw=ACMDarkBlue, thick,restrict x to domain=4:16] table [y index= 2,x index=0] {res-armsve/N1_N_heap_bench_res_1657156478.txt};

        	\addplot [nodes near coords, draw=ACMDarkBlue, thick,  mark=square, mark repeat={8},point meta=explicit, restrict x to domain=16:128] table [y index= 2,meta index=3, x index=0] {res-armsve/N1_N_heap_bench_res_1657156478.txt};
        	\end{axis}
        \end{tikzpicture}
        \subcaption{(N,1) $\rightarrow$ N}
        \label{fig:res-n1n}
    \end{subfigure}
    
    \begin{subfigure}[c]{0.5\textwidth}
        \begin{tikzpicture}
        	\begin{axis}[
        	    ymode=log,
        	    height=.2\textheight,
                width=\textwidth,
                axis lines*=left,
                ylabel={Execution time (s)},
                 every node near coord/.style={font=\normalsize},
        	]
        	\addplot [draw=ACMOrange , dashed] table [y index= 1,x index=0] {res-avx/NN_1_heap_bench_res_1658625605.txt};
            
        	\addplot [nodes near coords, draw=ACMOrange, thick, mark=square, mark repeat={8},point meta=explicit] table [y index= 2,meta index=3, x index=0] {res-avx/NN_1_heap_bench_res_1658625605.txt};

        	\addplot [draw=ACMDarkBlue , dashed] table [y index= 1,x index=0] {res-armsve/NN_1_heap_bench_res_1657156478.txt};

        	\addplot [draw=ACMDarkBlue, thick,restrict x to domain=4:16] table [y index= 2,x index=0] {res-armsve/NN_1_heap_bench_res_1657156478.txt};

        	\addplot [nodes near coords, draw=ACMDarkBlue, thick,  mark=square, mark repeat={8},point meta=explicit, restrict x to domain=16:128] table [y index= 2,meta index=3, x index=0] {res-armsve/NN_1_heap_bench_res_1657156478.txt};
        	\end{axis}
        \end{tikzpicture}
        \subcaption{(N,N) $\rightarrow$ 1}
        \label{fig:res-nn1}
    \end{subfigure}
    \begin{subfigure}[c]{0.49\textwidth}
        \begin{tikzpicture}
        	\begin{axis}[
        	    ymode=log,
        	    height=.2\textheight,
                width=\textwidth,
                axis lines*=left,
                 every node near coord/.style={font=\normalsize},
        	]
        	\addplot [draw=ACMOrange , dashed] table [y index= 1,x index=0] {res-avx/NN_N_heap_bench_res_1658625605.txt};

        	\addplot [nodes near coords, draw=ACMOrange, thick, mark=square, mark repeat={8},point meta=explicit] table [y index= 2,meta index=3, x index=0] {res-avx/NN_N_heap_bench_res_1658625605.txt};

        	\addplot [draw=ACMDarkBlue , dashed] table [y index= 1,x index=0] {res-armsve/NN_N_heap_bench_res_1657156478.txt};

        	\addplot [draw=ACMDarkBlue, thick,restrict x to domain=4:16] table [y index= 2,x index=0] {res-armsve/NN_N_heap_bench_res_1657156478.txt};

        	\addplot [nodes near coords, draw=ACMDarkBlue, thick,  mark=square, mark repeat={8},point meta=explicit, restrict x to domain=16:128] table [y index= 2,meta index=3, x index=0] {res-armsve/NN_N_heap_bench_res_1657156478.txt};
        	\end{axis}
        \end{tikzpicture}
        \subcaption{(N,N) $\rightarrow$ N}
        \label{fig:res-nnn}
    \end{subfigure}
    
    \begin{subfigure}[c]{0.5\textwidth}
        \begin{tikzpicture}
        	\begin{axis}[
        	    ymode=log,
        	    height=.2\textheight,
                width=\textwidth,
                axis lines*=left,
                ylabel={Execution time (s)},
                 every node near coord/.style={font=\normalsize},
        	]
        	\addplot [draw=ACMOrange , dashed] table [y index= 1,x index=0] {res-avx/NN_rN_heap_bench_res_1658625605.txt};
            
        	\addplot [nodes near coords, draw=ACMOrange, thick, mark=square, mark repeat={8},point meta=explicit] table [y index= 2,meta index=3, x index=0] {res-avx/NN_rN_heap_bench_res_1658625605.txt};

        	\addplot [draw=ACMDarkBlue , dashed] table [y index= 1,x index=0] {res-armsve/NN_rN_heap_bench_res_1657156478.txt};

        	\addplot [draw=ACMDarkBlue, thick,restrict x to domain=4:16] table [y index= 2,x index=0] {res-armsve/NN_rN_heap_bench_res_1657156478.txt};

        	\addplot [nodes near coords, draw=ACMDarkBlue, thick, mark=square, mark repeat={8},point meta=explicit, restrict x to domain=16:128] table [y index= 2,meta index=3, x index=0] {res-armsve/NN_rN_heap_bench_res_1657156478.txt};
        	\end{axis}
        \end{tikzpicture}
        \subcaption{(N,N) $\rightarrow$ r(N)}
        \label{fig:res-nnrn}
    \end{subfigure}
    \begin{subfigure}[c]{0.49\textwidth}
        \begin{tikzpicture}
        	\begin{axis}[
        	    ymode=log,
        	    height=.2\textheight,
                width=\textwidth,
                axis lines*=left,
                 every node near coord/.style={font=\normalsize},
        	]
        	\addplot [draw=ACMOrange , dashed] table [y index= 1,x index=0] {res-avx/rN1_1_heap_bench_res_1658625605.txt};

        	\addplot [nodes near coords, draw=ACMOrange, thick, mark=square, mark repeat={8},point meta=explicit] table [y index= 2,meta index=3, x index=0] {res-avx/rN1_1_heap_bench_res_1658625605.txt};

        	\addplot [draw=ACMDarkBlue , dashed] table [y index= 1,x index=0] {res-armsve/rN1_1_heap_bench_res_1657156478.txt};

        	\addplot [draw=ACMDarkBlue, thick,restrict x to domain=4:16] table [y index= 2,x index=0] {res-armsve/rN1_1_heap_bench_res_1657156478.txt};

        	\addplot [nodes near coords, draw=ACMDarkBlue, thick,  mark=square, mark repeat={8},point meta=explicit, restrict x to domain=16:128] table [y index= 2,meta index=3, x index=0] {res-armsve/rN1_1_heap_bench_res_1657156478.txt};
        	\end{axis}
        \end{tikzpicture}
        \subcaption{(r(N),1) $\rightarrow$ 1}
        \label{fig:res-rn11}
    \end{subfigure}
    
    \begin{subfigure}[c]{0.5\textwidth}
        \begin{tikzpicture}
        	\begin{axis}[
        	    ymode=log,
        	    height=.2\textheight,
                width=\textwidth,
                axis lines*=left,
                ylabel={Execution time (s)},
                 every node near coord/.style={font=\normalsize},
        	]
        	\addplot [draw=ACMOrange , dashed] table [y index= 1,x index=0] {res-avx/rN1_N_heap_bench_res_1658625605.txt};

        	\addplot [nodes near coords, draw=ACMOrange, thick, mark=square, mark repeat={8},point meta=explicit] table [y index= 2,meta index=3, x index=0] {res-avx/rN1_N_heap_bench_res_1658625605.txt};

        	\addplot [draw=ACMDarkBlue , dashed] table [y index= 1,x index=0] {res-armsve/rN1_N_heap_bench_res_1657156478.txt};

        	\addplot [draw=ACMDarkBlue, thick,restrict x to domain=4:16] table [y index= 2,x index=0] {res-armsve/rN1_N_heap_bench_res_1657156478.txt};

        	\addplot [nodes near coords, draw=ACMDarkBlue, thick,  mark=square, mark repeat={8},point meta=explicit, restrict x to domain=16:128] table [y index= 2,meta index=3, x index=0] {res-armsve/rN1_N_heap_bench_res_1657156478.txt};
        	\end{axis}
        \end{tikzpicture}
        \subcaption{(r(N),1) $\rightarrow$ N}
        \label{fig:res-rn1n}
    \end{subfigure}
    \begin{subfigure}[c]{0.49\textwidth}
        \begin{tikzpicture}
        	\begin{axis}[
        	    ymode=log,
        	    height=.2\textheight,
                width=\textwidth,
                axis lines*=left,
                 every node near coord/.style={font=\normalsize},
        	]
        	\addplot [draw=ACMOrange , dashed] table [y index= 1,x index=0] {res-avx/rNN_1_heap_bench_res_1658625605.txt};

        	\addplot [nodes near coords, draw=ACMOrange, thick, mark=square, mark repeat={8},point meta=explicit] table [y index= 2,meta index=3, x index=0] {res-avx/rNN_1_heap_bench_res_1658625605.txt};

        	\addplot [draw=ACMDarkBlue , dashed] table [y index= 1,x index=0] {res-armsve/rNN_1_heap_bench_res_1657156478.txt};

        	\addplot [draw=ACMDarkBlue, thick,restrict x to domain=4:16] table [y index= 2,x index=0] {res-armsve/rNN_1_heap_bench_res_1657156478.txt};

        	\addplot [nodes near coords, draw=ACMDarkBlue, thick,  mark=square, mark repeat={8},point meta=explicit, restrict x to domain=16:128] table [y index= 2,meta index=3, x index=0] {res-armsve/rNN_1_heap_bench_res_1657156478.txt};
        	\end{axis}
        \end{tikzpicture}
        \subcaption{(r(N),N) $\rightarrow$ 1}
        \label{fig:res-rnn1}
    \end{subfigure}
    
    \begin{subfigure}[c]{0.5\textwidth}
        \begin{tikzpicture}
        	\begin{axis}[
        	    ymode=log,
        	    height=.2\textheight,
                width=\textwidth,
                axis lines*=left,
                ylabel={Execution time (s)},
                xlabel={Test case size},
                 every node near coord/.style={font=\normalsize},
        	]
        	\addplot [draw=ACMOrange , dashed] table [y index= 1,x index=0] {res-avx/rNN_N_heap_bench_res_1658625605.txt};

        	\addplot [nodes near coords, draw=ACMOrange, thick, mark=square, mark repeat={8},point meta=explicit] table [y index= 2,meta index=3, x index=0] {res-avx/rNN_N_heap_bench_res_1658625605.txt};

        	\addplot [draw=ACMDarkBlue , dashed] table [y index= 1,x index=0] {res-armsve/rNN_N_heap_bench_res_1657156478.txt};

        	\addplot [draw=ACMDarkBlue, thick,restrict x to domain=4:16] table [y index= 2,x index=0] {res-armsve/rNN_N_heap_bench_res_1657156478.txt};

        	\addplot [nodes near coords, draw=ACMDarkBlue, thick,  mark=square, mark repeat={8},point meta=explicit, restrict x to domain=16:128] table [y index= 2,meta index=3, x index=0] {res-armsve/rNN_N_heap_bench_res_1657156478.txt};
        	\end{axis}
        \end{tikzpicture}
        \subcaption{(r(N),N) $\rightarrow$ N}
        \label{fig:res-rnnn}
    \end{subfigure}
    \begin{subfigure}[c]{0.49\textwidth}
        \begin{tikzpicture}
        	\begin{axis}[
        	    ymode=log,
        	    height=.2\textheight,
                width=\textwidth,
                axis lines*=left,
                xlabel={Test case size},
                 every node near coord/.style={font=\normalsize},
        	]
        	\addplot [draw=ACMOrange , dashed] table [y index= 1,x index=0] {res-avx/sNsN_N_heap_bench_res_1658625605.txt};

        	\addplot [nodes near coords, draw=ACMOrange, thick, mark=square, mark repeat={8},point meta=explicit] table [y index= 2,meta index=3, x index=0] {res-avx/sNsN_N_heap_bench_res_1658625605.txt};

        	\addplot [draw=ACMDarkBlue , dashed] table [y index= 1,x index=0] {res-armsve/sNsN_N_heap_bench_res_1657156478.txt};

        	\addplot [draw=ACMDarkBlue, thick,restrict x to domain=4:16] table [y index= 2,x index=0] {res-armsve/sNsN_N_heap_bench_res_1657156478.txt};

        	\addplot [nodes near coords, draw=ACMDarkBlue, thick, mark = square, mark repeat={8},point meta=explicit, restrict x to domain=16:128] table [y index= 2,meta index=3, x index=0] {res-armsve/sNsN_N_heap_bench_res_1657156478.txt};
        	\end{axis}
        \end{tikzpicture}
        \subcaption{(s(N),s(N)) $\rightarrow$ N}
        \label{fig:res-snsnn}
    \end{subfigure}
    \caption{Execution times for the different kernels and different problem sizes for the INTEL-AVX512 and ARM-SVE configurations.
    The speedup of the vectorized kernels by Autovesk against the automatic vectorization techniques of GCC compiler on scalar kernels is shown for some points.}
    \label{fig:res-speedup}
\end{figure}

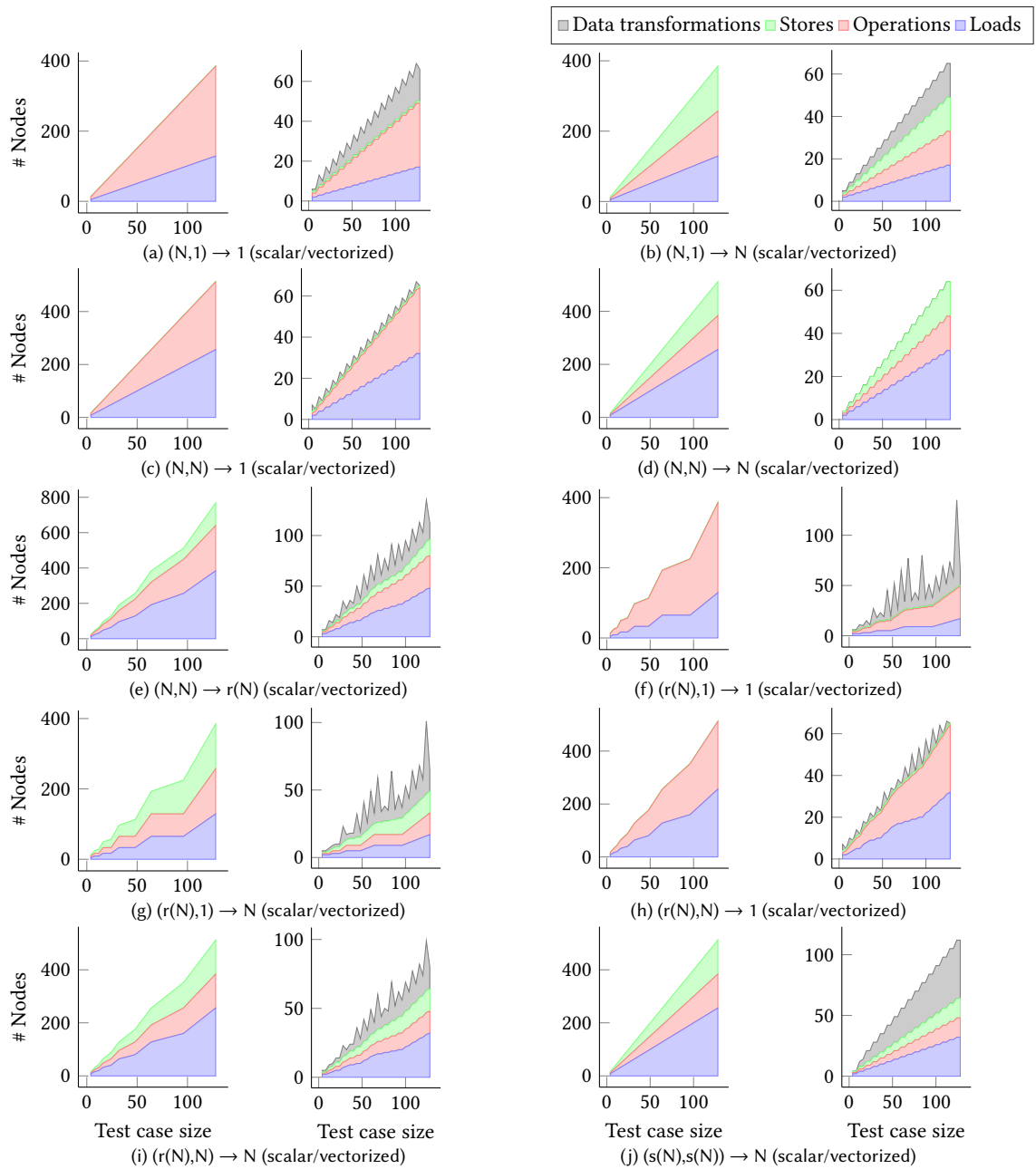
\begin{figure}[h!]
    \centering
    \hfill
        \begin{tikzpicture}
        \begin{axis}[%
                width=.5\textwidth,
        hide axis,
        xmin=10,
        xmax=50,
        ymin=0,
        ymax=0.4,
        legend columns=4,
        legend style={draw=white!15!black,legend cell align=left},
        legend style={at={(0.8,0.)}}
        ]
        \addlegendimage{draw=black!50, fill=black!20, mark=square*, only marks}
        \addlegendentry{Data transformations};
        \addlegendimage{draw=green!50, fill=green!20, mark=square*, only marks}
        \addlegendentry{Stores};
        \addlegendimage{draw=red!50, fill=red!20, mark=square*, only marks}
        \addlegendentry{Operations};
        \addlegendimage{draw=blue!50, fill=blue!20, mark=square*, only marks}
        \addlegendentry{Loads};
        \end{axis}
        \end{tikzpicture}
    
    \begin{subfigure}[c]{0.25\textwidth}
        \begin{tikzpicture}
        	\begin{axis}[
        	    height=.2\textheight,
                width=\textwidth,
                axis lines*=left,
                ylabel={\# Nodes},
		stack plots=y,
		area style,
        	]
        	\addplot [draw=blue!50, fill=blue!20] table [y index= 1,x index=0] {stats/N1_1_stats1657156478.txt}
		\closedcycle;
        	\addplot [draw=red!50, fill=red!20] table [y index=3,x index=0] {stats/N1_1_stats1657156478.txt}
		\closedcycle;
        	\addplot [draw=green!50, fill=green!20] table [y index=2,x index=0] {stats/N1_1_stats1657156478.txt}
		\closedcycle;
        	\end{axis}
        \end{tikzpicture}
    \end{subfigure}
    \begin{subfigure}[c]{0.23\textwidth}
        \begin{tikzpicture}
        	\begin{axis}[
        	    height=.2\textheight,
                width=\textwidth,
                axis lines*=left,
		stack plots=y,
		area style,
        	]
        	\addplot [draw=blue!50, fill=blue!20] table [y index= 5,x index=0] {stats/N1_1_stats1657156478.txt}
		\closedcycle;
        	\addplot [draw=red!50, fill=red!20] table [y index=7,x index=0] {stats/N1_1_stats1657156478.txt}
		\closedcycle;
        	\addplot [draw=green!50, fill=green!20] table [y index= 6,x index=0] {stats/N1_1_stats1657156478.txt}
		\closedcycle;
        	\addplot [draw=black!50, fill=black!20] table [y index= 8,x index=0] {stats/N1_1_stats1657156478.txt}
		\closedcycle;
        	\end{axis}
        \end{tikzpicture}
    \end{subfigure}
    \hfill
    \begin{subfigure}[c]{0.23\textwidth}
        \begin{tikzpicture}
        	\begin{axis}[
        	    height=.2\textheight,
                width=\textwidth,
                axis lines*=left,
		stack plots=y,
		area style,
        	]
        	\addplot [draw=blue!50, fill=blue!20] table [y index= 1,x index=0] {stats/N1_N_stats1657156478.txt}
		\closedcycle;
        	\addplot [draw=red!50, fill=red!20] table [y index=3,x index=0] {stats/N1_N_stats1657156478.txt}
		\closedcycle;
        	\addplot [draw=green!50, fill=green!20] table [y index=2,x index=0] {stats/N1_N_stats1657156478.txt}
		\closedcycle;
        	\end{axis}
        \end{tikzpicture}
    \end{subfigure}
    \begin{subfigure}[c]{0.23\textwidth}
        \begin{tikzpicture}
        	\begin{axis}[
        	    height=.2\textheight,
                width=\textwidth,
                axis lines*=left,
		stack plots=y,
		area style,
        	]
        	\addplot [draw=blue!50, fill=blue!20] table [y index= 5,x index=0] {stats/N1_N_stats1657156478.txt}
		\closedcycle;
        	\addplot [draw=red!50, fill=red!20] table [y index=7,x index=0] {stats/N1_N_stats1657156478.txt}
		\closedcycle;
        	\addplot [draw=green!50, fill=green!20] table [y index= 6,x index=0] {stats/N1_N_stats1657156478.txt}
		\closedcycle;
        	\addplot [draw=black!50, fill=black!20] table [y index= 8,x index=0] {stats/N1_N_stats1657156478.txt}
		\closedcycle;
        	\end{axis}
        \end{tikzpicture}
    \end{subfigure}
    
    \begin{subfigure}[c]{0.48\textwidth}
        \subcaption{(N,1) $\rightarrow$ 1 (scalar/vectorized)}
        \label{fig:instr:res-n11}
    \end{subfigure}
    \begin{subfigure}[c]{0.48\textwidth}
        \subcaption{(N,1) $\rightarrow$ N (scalar/vectorized)}
        \label{fig:instr:res-n1n}
    \end{subfigure}
    
    \begin{subfigure}[c]{0.25\textwidth}
        \begin{tikzpicture}
        	\begin{axis}[
        	    height=.2\textheight,
                width=\textwidth,
                axis lines*=left,
                ylabel={\# Nodes},
		stack plots=y,
		area style,
        	]
        	\addplot [draw=blue!50, fill=blue!20] table [y index= 1,x index=0] {stats/NN_1_stats1657156478.txt}
		\closedcycle;
        	\addplot [draw=red!50, fill=red!20] table [y index=3,x index=0] {stats/NN_1_stats1657156478.txt}
		\closedcycle;
        	\addplot [draw=green!50, fill=green!20] table [y index=2,x index=0] {stats/NN_1_stats1657156478.txt}
		\closedcycle;
        	\end{axis}
        \end{tikzpicture}
    \end{subfigure}
    \begin{subfigure}[c]{0.23\textwidth}
        \begin{tikzpicture}
        	\begin{axis}[
        	    height=.2\textheight,
                width=\textwidth,
                axis lines*=left,
		stack plots=y,
		area style,
        	]
        	\addplot [draw=blue!50, fill=blue!20] table [y index= 5,x index=0] {stats/NN_1_stats1657156478.txt}
		\closedcycle;
        	\addplot [draw=red!50, fill=red!20] table [y index=7,x index=0] {stats/NN_1_stats1657156478.txt}
		\closedcycle;
        	\addplot [draw=green!50, fill=green!20] table [y index=6,x index=0] {stats/NN_1_stats1657156478.txt}
		\closedcycle;
        	\addplot [draw=black!50, fill=black!20] table [y index= 8,x index=0] {stats/NN_1_stats1657156478.txt}
		\closedcycle;
        	\end{axis}
        \end{tikzpicture}
    \end{subfigure}
    \hfill
    \begin{subfigure}[c]{0.23\textwidth}
        \begin{tikzpicture}
        	\begin{axis}[
        	    height=.2\textheight,
                width=\textwidth,
                axis lines*=left,
		stack plots=y,
		area style,
        	]
        	\addplot [draw=blue!50, fill=blue!20] table [y index= 1,x index=0] {stats/NN_N_stats1657156478.txt}
		\closedcycle;
        	\addplot [draw=red!50, fill=red!20] table [y index=3,x index=0] {stats/NN_N_stats1657156478.txt}
		\closedcycle;
        	\addplot [draw=green!50, fill=green!20] table [y index=2,x index=0] {stats/NN_N_stats1657156478.txt}
		\closedcycle;
        	\end{axis}
        \end{tikzpicture}
    \end{subfigure}
    \begin{subfigure}[c]{0.23\textwidth}
        \begin{tikzpicture}
        	\begin{axis}[
        	    height=.2\textheight,
                width=\textwidth,
                axis lines*=left,
		stack plots=y,
		area style,
        	]
        	\addplot [draw=blue!50, fill=blue!20] table [y index= 5,x index=0] {stats/NN_N_stats1657156478.txt}
		\closedcycle;
        	\addplot [draw=red!50, fill=red!20] table [y index=7,x index=0] {stats/NN_N_stats1657156478.txt}
		\closedcycle;
        	\addplot [draw=green!50, fill=green!20] table [y index=6,x index=0] {stats/NN_N_stats1657156478.txt}
		\closedcycle;
        	\addplot [draw=black!50, fill=black!20] table [y index= 8,x index=0] {stats/NN_N_stats1657156478.txt}
		\closedcycle;
        	\end{axis}
        \end{tikzpicture}
    \end{subfigure}
    
    \begin{subfigure}[c]{0.48\textwidth}
        \subcaption{(N,N) $\rightarrow$ 1 (scalar/vectorized)}
        \label{fig:instr:res-nn1}
    \end{subfigure}
    \begin{subfigure}[c]{0.48\textwidth}
        \subcaption{(N,N) $\rightarrow$ N (scalar/vectorized)}
        \label{fig:instr:res-nnn}
    \end{subfigure}
    
    \begin{subfigure}[c]{0.25\textwidth}
        \begin{tikzpicture}
        	\begin{axis}[
        	    height=.2\textheight,
                width=\textwidth,
                axis lines*=left,
                ylabel={\# Nodes},
		stack plots=y,
		area style,
        	]
        	\addplot [draw=blue!50, fill=blue!20] table [y index= 1,x index=0] {stats/NN_rN_stats1657156478.txt}
		\closedcycle;
        	\addplot [draw=red!50, fill=red!20] table [y index=3,x index=0] {stats/NN_rN_stats1657156478.txt}
		\closedcycle;
        	\addplot [draw=green!50, fill=green!20] table [y index=2,x index=0] {stats/NN_rN_stats1657156478.txt}
		\closedcycle;
        	\end{axis}
        \end{tikzpicture}
    \end{subfigure}
    \begin{subfigure}[c]{0.23\textwidth}
        \begin{tikzpicture}
        	\begin{axis}[
        	    height=.2\textheight,
                width=\textwidth,
                axis lines*=left,
		stack plots=y,
		area style,
        	]
        	\addplot [draw=blue!50, fill=blue!20] table [y index= 5,x index=0] {stats/NN_rN_stats1657156478.txt}
		\closedcycle;
        	\addplot [draw=red!50, fill=red!20] table [y index=7,x index=0] {stats/NN_rN_stats1657156478.txt}
		\closedcycle;
        	\addplot [draw=green!50, fill=green!20] table [y index=6,x index=0] {stats/NN_rN_stats1657156478.txt}
		\closedcycle;
        	\addplot [draw=black!50, fill=black!20] table [y index= 8,x index=0] {stats/NN_rN_stats1657156478.txt}
		\closedcycle;
        	\end{axis}
        \end{tikzpicture}
    \end{subfigure}
    \hfill
    \begin{subfigure}[c]{0.23\textwidth}
        \begin{tikzpicture}
        	\begin{axis}[
        	    height=.2\textheight,
                width=\textwidth,
                axis lines*=left,
		stack plots=y,
		area style,
        	]
        	\addplot [draw=blue!50, fill=blue!20] table [y index= 1,x index=0] {stats/rN1_1_stats1657156478.txt}
		\closedcycle;
        	\addplot [draw=red!50, fill=red!20] table [y index=3,x index=0] {stats/rN1_1_stats1657156478.txt}
		\closedcycle;
        	\addplot [draw=green!50, fill=green!20] table [y index=2,x index=0] {stats/rN1_1_stats1657156478.txt}
		\closedcycle;
        	\end{axis}
        \end{tikzpicture}
    \end{subfigure}
    \begin{subfigure}[c]{0.23\textwidth}
        \begin{tikzpicture}
        	\begin{axis}[
        	    height=.2\textheight,
                width=\textwidth,
                axis lines*=left,
		stack plots=y,
		area style,
        	]
        	\addplot [draw=blue!50, fill=blue!20] table [y index= 5,x index=0] {stats/rN1_1_stats1657156478.txt}
		\closedcycle;
        	\addplot [draw=red!50, fill=red!20] table [y index=7,x index=0] {stats/rN1_1_stats1657156478.txt}
		\closedcycle;
        	\addplot [draw=green!50, fill=green!20] table [y index=6,x index=0] {stats/rN1_1_stats1657156478.txt}
		\closedcycle;
        	\addplot [draw=black!50, fill=black!20] table [y index= 8,x index=0] {stats/rN1_1_stats1657156478.txt}
		\closedcycle;
        	\end{axis}
        \end{tikzpicture}
    \end{subfigure}
    
    \begin{subfigure}[c]{0.48\textwidth}
        \subcaption{(N,N) $\rightarrow$ r(N) (scalar/vectorized)}
        \label{fig:instr:res-nnrn}
    \end{subfigure}
    \begin{subfigure}[c]{0.48\textwidth}
        \subcaption{(r(N),1) $\rightarrow$ 1 (scalar/vectorized)}
        \label{fig:instr:res-rn11}
    \end{subfigure}
    
    \begin{subfigure}[c]{0.25\textwidth}
        \begin{tikzpicture}
        	\begin{axis}[
        	    height=.2\textheight,
                width=\textwidth,
                axis lines*=left,
                ylabel={\# Nodes},
		stack plots=y,
		area style,
        	]
        	\addplot [draw=blue!50, fill=blue!20] table [y index= 1,x index=0] {stats/rN1_N_stats1657156478.txt}
		\closedcycle;
        	\addplot [draw=red!50, fill=red!20] table [y index=3,x index=0] {stats/rN1_N_stats1657156478.txt}
		\closedcycle;
        	\addplot [draw=green!50, fill=green!20] table [y index=2,x index=0] {stats/rN1_N_stats1657156478.txt}
		\closedcycle;
        	\end{axis}
        \end{tikzpicture}
    \end{subfigure}
    \begin{subfigure}[c]{0.23\textwidth}
        \begin{tikzpicture}
        	\begin{axis}[
        	    height=.2\textheight,
                width=\textwidth,
                axis lines*=left,
		stack plots=y,
		area style,
        	]
        	\addplot [draw=blue!50, fill=blue!20] table [y index= 5,x index=0] {stats/rN1_N_stats1657156478.txt}
		\closedcycle;
        	\addplot [draw=red!50, fill=red!20] table [y index=7,x index=0] {stats/rN1_N_stats1657156478.txt}
		\closedcycle;
        	\addplot [draw=green!50, fill=green!20] table [y index=6,x index=0] {stats/rN1_N_stats1657156478.txt}
		\closedcycle;
        	\addplot [draw=black!50, fill=black!20] table [y index= 8,x index=0] {stats/rN1_N_stats1657156478.txt}
		\closedcycle;
        	\end{axis}
        \end{tikzpicture}
    \end{subfigure}
    \hfill
    \begin{subfigure}[c]{0.23\textwidth}
        \begin{tikzpicture}
        	\begin{axis}[
        	    height=.2\textheight,
                width=\textwidth,
                axis lines*=left,
		stack plots=y,
		area style,
        	]
        	\addplot [draw=blue!50, fill=blue!20] table [y index= 1,x index=0] {stats/rNN_1_stats1657156478.txt}
		\closedcycle;
        	\addplot [draw=red!50, fill=red!20] table [y index=3,x index=0] {stats/rNN_1_stats1657156478.txt}
		\closedcycle;
        	\addplot [draw=green!50, fill=green!20] table [y index=2,x index=0] {stats/rNN_1_stats1657156478.txt}
		\closedcycle;
        	\end{axis}
        \end{tikzpicture}
    \end{subfigure}
    \begin{subfigure}[c]{0.23\textwidth}
        \begin{tikzpicture}
        	\begin{axis}[
        	    height=.2\textheight,
                width=\textwidth,
                axis lines*=left,
		stack plots=y,
		area style,
        	]
        	\addplot [draw=blue!50, fill=blue!20] table [y index= 5,x index=0] {stats/rNN_1_stats1657156478.txt}
		\closedcycle;
        	\addplot [draw=red!50, fill=red!20] table [y index=7,x index=0] {stats/rNN_1_stats1657156478.txt}
		\closedcycle;
        	\addplot [draw=green!50, fill=green!20] table [y index=6,x index=0] {stats/rNN_1_stats1657156478.txt}
		\closedcycle;
        	\addplot [draw=black!50, fill=black!20] table [y index= 8,x index=0] {stats/rNN_1_stats1657156478.txt}
		\closedcycle;
        	\end{axis}
        \end{tikzpicture}
    \end{subfigure}
    
    \begin{subfigure}[c]{0.48\textwidth}
        \subcaption{(r(N),1) $\rightarrow$ N (scalar/vectorized)}
        \label{fig:instr:res-rn1n}
    \end{subfigure}
    \begin{subfigure}[c]{0.48\textwidth}
        \subcaption{(r(N),N) $\rightarrow$ 1 (scalar/vectorized)}
        \label{fig:instr:res-rnn1}
    \end{subfigure}
    
    \begin{subfigure}[c]{0.25\textwidth}
        \begin{tikzpicture}
        	\begin{axis}[
        	    height=.2\textheight,
                width=\textwidth,
                axis lines*=left,
                ylabel={\# Nodes},
		stack plots=y,
		area style,
                xlabel={Test case size},
        	]
        	\addplot [draw=blue!50, fill=blue!20] table [y index= 1,x index=0] {stats/rNN_N_stats1657156478.txt}
		\closedcycle;
        	\addplot [draw=red!50, fill=red!20] table [y index=3,x index=0] {stats/rNN_N_stats1657156478.txt}
		\closedcycle;
        	\addplot [draw=green!50, fill=green!20] table [y index=2,x index=0] {stats/rNN_N_stats1657156478.txt}
		\closedcycle;
        	\end{axis}
        \end{tikzpicture}
    \end{subfigure}
    \begin{subfigure}[c]{0.23\textwidth}
        \begin{tikzpicture}
        	\begin{axis}[
        	    height=.2\textheight,
                width=\textwidth,
                axis lines*=left,
		stack plots=y,
		area style,
                xlabel={Test case size},
        	]
        	\addplot [draw=blue!50, fill=blue!20] table [y index= 5,x index=0] {stats/rNN_N_stats1657156478.txt}
		\closedcycle;
        	\addplot [draw=red!50, fill=red!20] table [y index=7,x index=0] {stats/rNN_N_stats1657156478.txt}
		\closedcycle;
        	\addplot [draw=green!50, fill=green!20] table [y index=6,x index=0] {stats/rNN_N_stats1657156478.txt}
		\closedcycle;
        	\addplot [draw=black!50, fill=black!20] table [y index= 8,x index=0] {stats/rNN_N_stats1657156478.txt}
		\closedcycle;
        	\end{axis}
        \end{tikzpicture}
    \end{subfigure}
    \hfill
    \begin{subfigure}[c]{0.23\textwidth}
        \begin{tikzpicture}
        	\begin{axis}[
        	    height=.2\textheight,
                width=\textwidth,
                axis lines*=left,
		stack plots=y,
		area style,
                xlabel={Test case size},
        	]
        	\addplot [draw=blue!50, fill=blue!20] table [y index= 1,x index=0] {stats/sNsN_N_stats1657156478.txt}
		\closedcycle;
        	\addplot [draw=red!50, fill=red!20] table [y index=3,x index=0] {stats/sNsN_N_stats1657156478.txt}
		\closedcycle;
        	\addplot [draw=green!50, fill=green!20] table [y index=2,x index=0] {stats/sNsN_N_stats1657156478.txt}
		\closedcycle;
        	\end{axis}
        \end{tikzpicture}
    \end{subfigure}
    \begin{subfigure}[c]{0.23\textwidth}
        \begin{tikzpicture}
        	\begin{axis}[
        	    height=.2\textheight,
                width=\textwidth,
                axis lines*=left,
		stack plots=y,
		area style,
                xlabel={Test case size},
        	]
        	\addplot [draw=blue!50, fill=blue!20] table [y index= 5,x index=0] {stats/sNsN_N_stats1657156478.txt}
		\closedcycle;
        	\addplot [draw=red!50, fill=red!20] table [y index=7,x index=0] {stats/sNsN_N_stats1657156478.txt}
		\closedcycle;
        	\addplot [draw=green!50, fill=green!20] table [y index=6,x index=0] {stats/sNsN_N_stats1657156478.txt}
		\closedcycle;
        	\addplot [draw=black!50, fill=black!20] table [y index= 8,x index=0] {stats/sNsN_N_stats1657156478.txt}
		\closedcycle;
        	\end{axis}
        \end{tikzpicture}
    \end{subfigure}
    
    \begin{subfigure}[c]{0.48\textwidth}
        \subcaption{(r(N),N) $\rightarrow$ N (scalar/vectorized)}
        \label{fig:instr:res-rnnn}
    \end{subfigure}
    \begin{subfigure}[c]{0.48\textwidth}
        \subcaption{(s(N),s(N)) $\rightarrow$ N (scalar/vectorized)}
        \label{fig:instr:res-snsnn}
    \end{subfigure}
    
    \caption{Number and types of graph nodes for the different kernels and different problem sizes.
    The data transformation operations only exist for the vectorized kernels because they apply to vectors.
    The number of resulting binary instructions can differ but will be highly correlated to the numbers shown.}
    \label{fig:instr}
\end{figure}

Our performance study compared the automatically vectorized C++ code generated by Autovesk with the automatic vectorization techniques of scalar kernels produced by the GCC compiler, the Clang LLVM-based compiler, and the Intel icpx compiler.
Figure~\ref{fig:res-speedup} shows the execution times of the 10 kernels with different problem sizes for both the INTEL AVX512 and ARM-SVE configurations. 
The comparison is made between the vectorized code generated by Autovesk and the automatic vectorization of GCC compiler on scalar codes.
The results show that vectorized kernels provide a noticeable speedup over scalar kernels, as indicated by the speedup shown at some points.
In Figure~\ref{fig:instr}, we provide a detailed breakdown of the number and types of graph nodes for the different kernels and problem sizes in our study. 
There are scalar graph nodes, including loads, stores and operations, and vectorial graph nodes for the generated vectorized code, including loads, stores, operations and data transformations. 
Importantly, the data transformation operations exist only for the vectorized kernels, as they apply to vectors. 
Although the number of resulting binary instructions may differ, it is highly correlated with the number of graph nodes shown. 
These results provide insights into the complexity of the vectorization process and the types of operations involved in generating efficient vectorized code from scalar code.

From the execution times, we observe that our approach provides a significant speedup for all kernels on both configurations, except for kernels $(N,1) \rightarrow N$ and  $(N,N) \rightarrow N$.
For these kernels, the compiler is able to vectorize, such that our approach obtains very similar execution times.
The kernels $(N,1) \rightarrow 1$ and $(N,N) \rightarrow 1$ can also be vectorized easily, but it requires a reduction.
We can see, Figures~\ref{fig:res-n11} and~\ref{fig:res-nn1}, that our approach is faster for all sizes on ARM-SVE and after a fairly large size on INTEL-AVX512.
When using random access ($r(N)$) or a shift ($s(N)$) our method can provide a significant speedup of 12.
The execution times then can vary and we can understand this behavior by looking at Figure~\ref{fig:instr}.
For instance, for the kernel $(r(N),1) \rightarrow 1$, execution time Figure~\ref{fig:res-rn11} and operations Figure~\ref{fig:instr:res-rn11}, we can see that the data transformation operations vary significantly depending on the size.
This is because we do not always use a multiple of $8$ (the length of the SIMD vector), which leads to different best instruction graphs.
This effect is visible in Figures~\ref{fig:instr:res-n11},~\ref{fig:instr:res-rn1n} and~\ref{fig:instr:res-nn1} but with less impact on the performance.
As expected for the kernel $(N,N) \rightarrow N$, Figure~\ref{fig:instr:res-nnn}, there are no data transformation operations.
Among the strategies to fix the order of the vector’s elements, 76\% of the best configurations were obtained by leaving the elements in their original order, which is not surprising due to the types of kernels we test.
Then, 16\% were obtained with the partitioning strategy and 8\% with the clustering strategy.

Figure~\ref{clang:icpx:speedups} shows the speedups achieved by Autovesk compared to the Intel icpx and Clang compilers for the 10 kernels selected in our study.
To evaluate performance, we ran each kernel as we did for the comparison with GCC, i.e.\ with different problem sizes, but instead of showing all the details, we calculated the average speedup along with the minimum and maximum variation over the automatic vectorization techniques of the compilers.
The evaluation showed that the performance of the vectorized kernels using Autovesk is above the compiler versions for most kernels.
For the cases "(N,N) $\rightarrow$ N" and "(N,1) $\rightarrow$ N" using Intel icpx, see~Figure~\ref{fig_speedup_intel}, we obtain similar performance, which is what we expect since the data accesses are contiguous and linear.
However, Clang did not deliver the same performance, see Figure~\ref{fig_speedup_clang}. 

\pgfplotsset{every error bar/.style={line width=0.5mm}}

\pgfplotstableread{
x        y  y-min   y-max
sNsN\_N 3.82 2.97356 5.12179
rNN\_1 4.30664 3.11583 5.57086
rN1\_N 2.80845 2.08232 6.21249
NN\_rN 4.37212 2.86589 4.52856
rNN\_N 2.06565 1.49373 3.56722
rN1\_1 4.15309 3.57144 6.08301
NN\_1 4.54015 3.7033 6.22435
NN\_N 1.89107 1.30741 2.30152
N1\_N 2.29013 1.46667 3.50443
N1\_1 4.07319 3.15221 5.82327
}{\clangMinAvgMaxHeap}

\pgfplotstableread{
x        y  y-min   y-max
sNsN\_N 3.08974 2.64614 3.58465
rNN\_1 6.32493 5.35307 3.3429
rN1\_N 2.63294 2.35536 8.5519
NN\_rN 4.54291 4.05608 5.09126
rNN\_N 2.08865 1.84881 4.76054
rN1\_1 7.64182 7.13855 9.30169
NN\_1 2.674 1.74449 4.18368
NN\_N 1.03843 0.400634 0.407095
N1\_N 1.15076 0.398643 0.357189
N1\_1 2.35674 1.48574 4.24605
}{\icpxMinAvgMaxHeap}

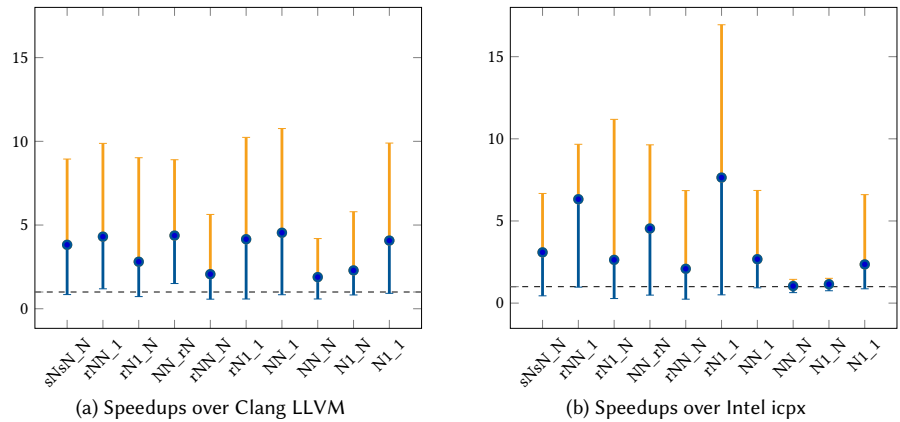
\begin{figure}[h!]
    \centering
    \subfloat[Speedups over Clang LLVM]{\label{fig_speedup_clang}
        \begin{tikzpicture}[scale=0.75] 
        \begin{axis} [
            ymax=18,
            xticklabel style={rotate=45},
            symbolic x coords={sNsN\_N, rNN\_1, rN1\_N, NN\_rN, rNN\_N, rN1\_1, NN\_1, NN\_N, N1\_N, N1\_1},
            xtick=data,
             extra y ticks = 1,
             extra y tick labels={},
             extra y tick style={grid=major,major grid style={dashed,draw=black}}
        ]
        \addplot+[ACMOrange, very thick, forget plot,only marks] 
          plot[very thick, error bars/.cd, y dir=plus, y explicit]
          table[x=x,y error expr=\thisrow{y-max},y=y] {\clangMinAvgMaxHeap};
        \addplot+[ACMDarkBlue, very thick, only marks,xticklabels=\empty] 
          plot[very thick, error bars/.cd, y dir=minus, y explicit]
          table[x=x,y error expr=\thisrow{y-min},y=y] {\clangMinAvgMaxHeap};
        \end{axis} 
        \end{tikzpicture}
    }
    \qquad
    \subfloat[Speedups over Intel icpx]{\label{fig_speedup_intel}
        \begin{tikzpicture}[scale=0.75]
        \begin{axis} [
            ymax=18,
            xticklabel style={rotate=45},
            symbolic x coords={sNsN\_N, rNN\_1, rN1\_N, NN\_rN, rNN\_N, rN1\_1, NN\_1, NN\_N, N1\_N, N1\_1},
            xtick=data,
             extra y ticks = 1,
             extra y tick labels={},
             extra y tick style={grid=major,major grid style={dashed,draw=black}}
        ]
        \addplot+[ACMOrange, very thick, forget plot,only marks] 
          plot[very thick, error bars/.cd, y dir=plus, y explicit]
          table[x=x,y error expr=\thisrow{y-max},y=y] {\icpxMinAvgMaxHeap};
        \addplot+[ACMDarkBlue, very thick, only marks,xticklabels=\empty] 
          plot[very thick, error bars/.cd, y dir=minus, y explicit]
          table[x=x,y error expr=\thisrow{y-min},y=y] {\icpxMinAvgMaxHeap};
        \end{axis} 
        \end{tikzpicture}
    }
    \caption{Comparison of Autovesk's automatic vectorization speedups against the Clang LLVM-based and Intel icpx compilers on 10 selected kernels. Each kernel was executed with different problem sizes. The minimum, average and maximum speedup over the compilers' auto-vectorization techniques are shown.}
    \label{clang:icpx:speedups}
\end{figure}

\begin{figure}[h!]
    \centering
    \pgfplotstableread[col sep=comma]{extrakernels.csv}\mydata 
    \begin{tikzpicture}
        \begin{axis}[
        x tick label style={/pgf/number format/1000 sep={}},
        ylabel=Speedup,
        legend style={at={(0.5,-0.15)},anchor=north,legend columns=-1},            
        ybar,
        bar width=8pt,
        enlarge x limits=0.2,
        xtick=data,
        xticklabels from table={\mydata}{kernel},  
        nodes near coords={\pgfmathprintnumber[fixed,precision=4]{\pgfplotspointmeta}},
        every node near coord/.append style={font=\footnotesize, anchor=west, rotate=90},
        point meta=explicit,
        ymax=5
        ]
            \addplot[ACMBlue, fill=ACMBlue] table[x expr=\coordindex, y=Gcc, meta=Gcct]{\mydata};
            \addplot[ACMDarkBlue, fill=ACMDarkBlue] table[x expr=\coordindex, y=Gccafx, meta=Gccafxt]{\mydata};
            \addplot[ACMOrange, fill=ACMOrange] table[x expr=\coordindex, y=Clang, meta=Clangt]{\mydata};
            \addplot[ACMPurple, fill=ACMPurple] table[x expr=\coordindex, y=Icpx, meta=Icpxt]{\mydata};
            \legend{Gcc, Gcc-A64FX, Clang, icpx}
        \end{axis}
    \end{tikzpicture}
    \caption{Comparison of Autovesk's automatic vectorization speedup against the GNU compiler (GCC) on x86 architecture and A64FX 64-bit ARM architecture, Clang LLVM-based compiler and the Intel (icpx) compiler for 4 different numerical kernels.
    We provide the execution times of the Autovesk versions above each bar.}
    \label{fig:extra-kernels}
\end{figure}
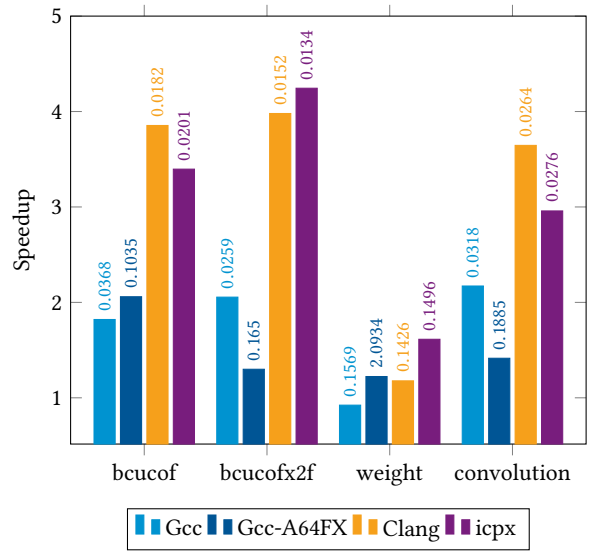

\paragraph{Set-NK}
Figure~\ref{fig:extra-kernels} provides the results for the numerical kernels.
Our study highlights the relative advantages of Autovesk over modern compilers in optimizing the numerical computations presented. 
In particular, Autovesk achieves significant speedups over Clang and Intel compilers, exceeding the speedups achieved over GCC.
For the "bcucof" kernel, Autovesk runs 1.8 times faster than GCC. 
It also outperforms Clang LLVM by 3.8x with an execution time of 0.0182. 
Autovesk achieves a remarkable 2x speedup over GCC for the "bcucofx2f" kernel. Compared to Intel icpx, Autovesk achieves a significant speedup of 4.2x, reducing the execution time to 0.0134. 
Autovesk performs well on the "weight" kernel and excels on the "convolution" kernel with speedups of 2.1x over GCC, 3.6x over LLVM and 2.9x over Intel icpx, reducing the execution time to 0.0276. 
These consistent and significant performance improvements highlight the effectiveness of Autovesk in optimizing numerical kernels. 
Further investigation is required to explore its potential in broader application domains.

\section{Stack/register exchange minimization}
To alleviate the pressure on the registers and minimize stack/register exchange, we proposed a reordering algorithm for the generated C++ code in Section~\ref{sec:back}.

\subsection{Test cases}

To evaluate the effectiveness of our proposed reordering algorithm, we generated random instructions using our test case generator, Pred\emph{X}. 
This generator creates random instructions that have dependencies between variables. 
The \emph{X} in Pred\emph{X} represents, for a given variable, the maximum number of predecessors. 
To generate the test cases, we start with a single variable and link it to previous variables with a jump between 1 and 10, meaning that the first 9 variables could have zero predecessors. 
A variable with zero predecessors is considered a load and a variable with zero successors is considered a store. 
We generated AVX512 code, where each variable is an AVX512 vector, and simply added the input. 
This allowed us to test the performance of our reordering algorithm on randomly generated code of varying complexity.
Our test case generator is referred to as Pred\emph{X}, where \emph{X} means that a variable is computed using from 1 to \emph{X} variables (i.e. in the dependency graph, a variable can have up to \emph{X} predecessors).

\subsection{Results}

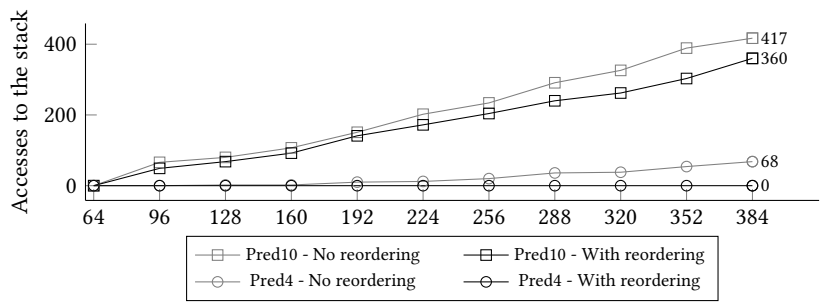
\begin{figure}[h!]
    \begin{tikzpicture}
    	\begin{axis}[
    	    height=.2\textheight,
            width=.75\textwidth,
            xmin=60,
            axis lines*=left,
            legend style={
                at={(0.5,-0.2)}, 
                anchor=north,
                legend columns=2,
                font=\footnotesize,
                /tikz/every even column/.append style={column sep=0.5cm},
            },
            ylabel={Accesses to the stack},
            xlabel={Test case size},
            xtick=data,
            xticklabels from table={res-min-registers.txt}{Size},
            every node near coord/.style={font=\footnotesize},
            nodes near coords align={right},
    	]
    	\addplot [draw=black!50, mark=square] table [y index= 6,x index=0] {res-min-registers.txt};
        \addlegendentry{Pred10 - No reordering}
    	\addplot [draw=black, mark=square] table [y index= 5,x index=0] {res-min-registers.txt};
        \addlegendentry{Pred10 - With reordering}
        
    	\addplot [draw=black!50, mark=o] table [y index= 2,x index=0] {res-min-registers.txt};
        \addlegendentry{Pred4 - No reordering}
    	\addplot [draw=black, mark=o] table [y index= 1,x index=0] {res-min-registers.txt};
        \addlegendentry{Pred4 - With reordering}

    	\addplot [nodes near coords, draw=black!50, mark=square] table [y index= 6,x index=0] {res-min-registers-last.txt};
    	\addplot [nodes near coords, draw=black, mark=square] table [y index= 5,x index=0] {res-min-registers-last.txt};
    	\addplot [nodes near coords, draw=black!50, mark=o] table [y index= 2,x index=0] {res-min-registers-last.txt};
    	\addplot [nodes near coords, draw=black, mark=o] table [y index= 1,x index=0] {res-min-registers-last.txt};
    	\end{axis}
    \end{tikzpicture}
    \caption{Number of accesses to the stack (push, pop or direct accesses) for different test case sizes.
    Pred\emph{X} refers to a test case where a new variable is constructed using up to X other variables as input.}
    \label{fig:res-min-registers}
\end{figure}

Using our test case generator Pred\emph{X}, we evaluated the effectiveness of our proposed reordering algorithm by generating random instructions for two different test case configurations: Pred\emph{4} and Pred\emph{10}, while varying the problem sizes. 
We provide the results in Figure~\ref{fig:res-min-registers}, showing the number of accesses to the stack (push, pop or direct accesses) with and without the reordering algorithm.
We can observe that reordering the instructions with our strategy always provide a benefit.
More precisely, for Pred\emph{4}, the compiler can avoid using the stack for all problem sizes but will do 68 registers spinning with the original order.
For Pred\emph{10}, the gain is around 15\%.

\section{Limitations and future works}
Our research has focused on automatically vectorizing unstructured static kernels with irregular data access patterns, but there are several limitations and opportunities for future improvement.
One area we intend to address is extending the applicability of Autovesk to cases where non-static loops can be expressed as repetitions of static ones.
Indeed, non-static kernels can be divided into those that can be described by the polyhedral model, for which our approach is not beneficial and those that cannot. 
For the latter, there are non-static kernels that require alternative optimization methods, and those that can be viewed as repetitions of static kernels, where we aim to apply our method.
Although our tool automatically generates vectorized code, at this stage, it still requires the user to provide input code in a specific format, as we are using C++ templates to extract the graph of instructions. 
To eliminate user intervention and extend the range of codes that can be automatically vectorized, we plan to integrate Autovesk's transformations into an existing compiler. 
In addition, we can investigate incorporating a cost model for each strategy in the graph transformations. 
By considering the potential benefits and overheads associated with vectorization, we can make informed decisions and avoid unnecessary vectorization where it may not provide significant performance gains.
These directions aim to improve Autovesk's usability and effectiveness in more applications.

\section{Conclusion}

We introduced Autovesk, an automatic vectorization tool designed for complex computational kernels.
We presented our strategy that relies on heuristic-based algorithms to avoid paying the price of finding an optimal solution.
Our method generates several solutions and returns the one with the lowest number of operations.
To assess the vectorization performance of Autovesk, we a comparative evaluation against three leading compilers: GCC, LLVM, and Intel, using a set of proposed kernels, including both custom and real-world numerical kernels.
We demonstrate that Autovesk can provide a significant speedup on several kernels.
It remains competitive even in cases where general compilers can vectorize.
Furthermore, Autovesk includes a procedure to reorder the instructions to reduce register spilling.
While Autovesk shows promising results for the automatic vectorization of unstructured static kernels, there is still room for improvement and extension. 
Our ongoing research will continue to explore these possibilities.

\begin{acks}
This work used the Isambard 2 UK National Tier-2 HPC Service (\url{http://gw4.ac.uk/isambard/}) operated by GW4 and the UK Met Office, which is an EPSRC project (EP/T022078/1).
We also used the PlaFRIM experimental testbed, supported by Inria, CNRS (LABRI and IMB), Université de Bordeaux, Bordeaux INP and Conseil Régional d’Aquitaine (\url{https://www.plafrim.fr}). 
\end{acks}

\bibliographystyle{ACM-Reference-Format}
\bibliography{autovesk}

\FloatBarrier
\appendix

\section{Partitioning strategy}
\label{app:partitioning}
\begin{algorithm}[H]
\footnotesize
 \caption{Converting a group into a set of $V$ sub-groups, where each group has at most $VEC\_SIZE$ elements.}
 \label{algo:partitioning}
    \tcp{Given a list of elements $e$ and a matrix score $d$}
    \tcp{We start with one partition containing all the elements}
    wip\_partitions.push(new\_partition(e)) \;
    final\_partitions $\gets \emptyset$ \;
    
    \While{|final\_partitions| $\neq V$}{ |\label{algo:partitioning:s2}|
        $p \gets$ wip\_partitions.pop() \;
        \If{$|p| \leq VEC\_SIZE$}{
            final\_partitions.insert($p$)\;
            continue;
        }
        \tcp{We compute the desired partition sizes}
        vecs\_in\_p $\gets (|p|+VEC\_SIZE-1)/VEC\_SIZE$\;
        margin $\gets$ vecs\_in\_p $\times VEC\_SIZE - |p|$\;
        size\_max\_p1 $\gets (|p|/2) \times VEC\_SIZE$ \;
        size\_max\_p2 $\gets (|p|/2) \times VEC\_SIZE$ \;
        size\_min\_p1 $\gets$ size\_max\_p1 - margin \;
        size\_min\_p2 $\gets$ size\_max\_p2 - margin \;
        \tcp{Find the elements in p with the lowest score}
        |\label{algo:partitioning:s3}|
        $(i,j) \gets$ find\_min\_score(p , d)\;
        |\label{algo:partitioning:s4}|
        sub\_p1 $\gets$ i\;
        sub\_p2 $\gets$ j\;
        p.remove(i, j)\;
        \While{$|p| \neq 0$}{
            \uIf{$|sub\_p1| == size\_max\_p1$}{
                sub\_p2.insert(p.pop())\;
            }
            \uElseIf{$|sub\_p2| == size\_max\_p2$}{
                sub\_p1.insert(p.pop())\;
            }
            \Else{
                \tcp{Find the element n that has the highest accumulated score either with sub\_p1 or sub\_p2}
                |\label{algo:partitioning:s5}|
                $[n, score1, score2] \gets$ find\_highest\_score(p, sub\_p1, sub\_p2)\;
                p.remove(n)\;
                \uIf{$score1 \geq score2$ OR ($score1 == score2$ AND $size\_max\_p1-|p1| \geq size\_max\_p2-|p2| $)}{
                    sub\_p1.insert(n)\;
                }
                \Else{
                    sub\_p2.insert(p.pop())\;
                }
            }
        }
        wip\_partition.insert(sub\_p1, sub\_p2)\;
    }
    
\end{algorithm}

\section{Clustering strategy}
\label{app:clustering}
\begin{algorithm}[H]
\footnotesize
 \caption{Converting a group into a set of $V$ sub-groups, where each group has at most $VEC\_SIZE$ elements.}
 \label{algo:clustering}
    \tcp{Given a list of elements $e$ and a matrix score $d$}
    \tcp{We initiate the sub-groups}
    $subgroups[V] \gets \emptyset$ \;
    |\label{algo:clustering:s1}|
    (i,j) $\gets$ find\_min(d)\; 
    e $\gets$ remove(e, i, j)\;
    subgroup[0].add(i)\;
    subgroup[1].add(j)\;
    
    |\label{algo:clustering:s2}|
    \For{v from 2 to $V$}{ 
        worst\_idx $\gets 0$\;
        worst\_score $\gets +\inf$\;
        \For{idx \textbf{in} $e$}{
            score $\gets \sum_{i=0}^{v}{d(subgroup[i][0], idx)}$\;
            \If{score < worst\_score}{
                score = worst\_score\;
                worst\_idx = idx\;
            }
        }
        e $\gets$ remove(e, worst\_idx)\;
        subgroup[v].add(worst\_idx)\;
    }
    
    |\label{algo:clustering:s3}|
    \While{e is not empty}{ 
        best\_idx $\gets 0$\;
        best\_subgroup $\gets 0$\;
        best\_score $\gets -\inf$\;
        \For{v from 0 to $V$}{
            \For{idx \textbf{in} $e$}{
                score $\gets \sum_{i=0}^{v}{d(subgroup[i][:], idx)}$\;
                \If{score > best\_score}{
                    score = best\_score\;
                    best\_idx = idx\;
                    best\_subgroup = v\;
                }
            }
        }
        e $\gets$ remove(e, best\_idx)\;
        subgroup[best\_subgroup].add(best\_idx)\;
    }    
\end{algorithm}

\end{document}